%% file: main.tex
\documentclass{article}

\usepackage{algorithm}
\usepackage[margin=30mm,includehead,includefoot]{geometry}
\usepackage[noend]{algpseudocode}
\usepackage{amsmath,amssymb,amsthm,amscd,color,comment}
\usepackage{mathtools}
\usepackage{autobreak}
\usepackage{color}
\usepackage{xcolor}
\usepackage{url}
\usepackage{blkarray}
\usepackage{setspace}
\usepackage{jheppub}
\makeatletter
\def\@fpheader{$\quad$}
\makeatother
\usepackage{empheq}
\usepackage{graphicx}
\usepackage{booktabs, multirow}
\usepackage{subcaption}
\usepackage{bm}
\usepackage{enumerate}
\usepackage{comment}
\usepackage{parskip}
\usepackage{bbm}
\usepackage{slashed}
\usepackage{colortbl}
\usepackage{nicematrix}

\usepackage{mathbbol}
\usepackage{float}
\usepackage{booktabs}
\usepackage{tikz}

\usepackage{listings}
\usepackage{xcolor}
\usepackage{mdframed}

\usepackage{cleveref}

\definecolor{light-gray}{gray}{0.95}

\definecolor{gh-fg}{HTML}{000000}
\definecolor{gh-comment}{HTML}{007ACC}
\definecolor{gh-keyword}{HTML}{E60000}
\definecolor{gh-func}{HTML}{FF7F00}
\definecolor{gh-string}{HTML}{999999} 
\definecolor{gh-number}{HTML}{000000}
\definecolor{gh-linenum}{HTML}{666666}
\definecolor{gh-frame}{HTML}{BBBBBB}

\lstnewenvironment{fullpagecode}[1][]%
  {\noindent\begin{minipage}{\textwidth}\begin{lstlisting}[#1]}%
  {\end{lstlisting}\end{minipage}}
\lstdefinestyle{githubMathematica}{
  language=Mathematica,
  backgroundcolor=\color{light-gray},
  basicstyle={\small\setstretch{1}\def\fvm@Scale{0.85}\ttfamily\selectfont\color{gh-fg}},
  commentstyle=\itshape\color{gh-comment},
  keywordstyle=\bfseries\color{gh-fg},
  stringstyle=\color{gh-string},
  identifierstyle=\color{gh-fg},
  numbers=right,
  numberstyle={\scriptsize\sffamily\color{gh-linenum}},
  numbersep=8pt,
  frame=lines,
  rulecolor=\color{gh-frame},
  tabsize=4,
  showstringspaces=false,
  breaklines=false,
  numbers=none,
  columns=fixed,
  basewidth=0.5em,
  aboveskip=1em,
  belowskip=0pt,
  emph={FindIrreducibleMonomials,BuildPolynomialSystem,
        ReconstructPolynomialRemainder,BuildCompanionMatrices,
        BuildTargetCompanionMatrices,ReconstructTargetCompanionMatrices,
        BuildCharacteristicPolynomials,ReconstructCharacteristicPolynomials,
        FFDet,SortVariables,FindEliminationMonomials,BuildEliminationSystems,ReconstructEliminationSystems},
  emphstyle=\bfseries\color{purple},  %
  literate=
    *{0}{{{\color{gh-number}0}}}1 {1}{{{\color{gh-number}1}}}1
     {2}{{{\color{gh-number}2}}}1 {3}{{{\color{gh-number}3}}}1
     {4}{{{\color{gh-number}4}}}1 {5}{{{\color{gh-number}5}}}1
     {6}{{{\color{gh-number}6}}}1 {7}{{{\color{gh-number}7}}}1
     {8}{{{\color{gh-number}8}}}1 {9}{{{\color{gh-number}9}}}1
     {.0}{{{\color{gh-number}.0}}}2 {.1}{{{\color{gh-number}.1}}}2
     {.2}{{{\color{gh-number}.2}}}2 {.3}{{{\color{gh-number}.3}}}2
     {.4}{{{\color{gh-number}.4}}}2 {.5}{{{\color{gh-number}.5}}}2
     {.6}{{{\color{gh-number}.6}}}2 {.7}{{{\color{gh-number}.7}}}2
     {.8}{{{\color{gh-number}.8}}}2 {.9}{{{\color{gh-number}.9}}}2
}
\newcommand{\soft}[1]{\textsc{#1}}

\newcommand{\code}[1]{\texttt{#1}}

\makeatletter
\newlength{\apb@width}
\newcommand{\autoparbox}[2][c]{
    \settowidth{\apb@width}{#2}%
    \parbox[#1]{\apb@width}{#2}%
}
\newcommand{\includegraphicsbox}[2][]{\autoparbox{\includegraphics[#1]{#2}}}
\makeatother

\definecolor{green1}{HTML}{3D792A}
\definecolor{cyan1}{HTML}{37cdaa}
\definecolor{blue1}{HTML}{5d7ac4}
\definecolor{purple1}{HTML}{845ea8}
\definecolor{orange1}{HTML}{e07229}
\definecolor{yellow1}{HTML}{edcb52}
\definecolor{red}{HTML}{921818}
\colorlet{red1}{red!80!white}
\colorlet{red2}{red!40!white}
\colorlet{red3}{red!20!white}
\definecolor{purple}{HTML}{630330}
\colorlet{purple1}{purple!80!white}
\colorlet{purple2}{purple!40!white}
\definecolor{yellow}{HTML}{f4e097}
\definecolor{gr}{gray}{0.7}
\definecolor{gr9}{gray}{0.9}
\definecolor{gr8}{gray}{0.8}
\definecolor{gr7}{gray}{0.7}
\definecolor{gr6}{gray}{0.6}
\definecolor{gr5}{gray}{0.5}
\definecolor{gr4}{gray}{0.4}
\definecolor{gr3}{gray}{0.3}
\definecolor{gr2}{gray}{0.2}
\definecolor{gr1}{gray}{0.1}
\newcommand{\gr}[1]{{\color{gr}#1}}

\newcommand{\brk}[1]{(#1)}

\newcommand{\brc}[1]{\{#1\}}

\newcommand{\bigbrc}[1]{\bigl\{#1\bigr\}}
\newcommand{\Bigbrc}[1]{\Bigl\{#1\Bigr\}}
\newcommand{\abs}[1]{|#1|}
\newcommand{\vev}[1]{\langle #1\rangle}

\newcommand{\supbrk}[1]{^{\brk{#1}}}

\newcommand{\dd}{\mathrm{d}}
\newcommand{\Integers}{\mathbb{Z}}
\newcommand{\Rationals}{\mathbb{Q}}
\newcommand{\Complex}{\mathbb{C}}

\newcommand{\mId}{\mathbb{1}}
\newcommand{\mzero}{\gr{\cdot}}
\newcommand{\mzeroRed}{\color{red!25}{\cdot}}
\newcommand{\mzeroPurple}{\color{purple!25}{\cdot}}
\newcommand{\mZero}{\gr{0}}
\newcommand{\mons}{\mathbf{m}}
\newcommand{\Ideal}{\mathcal{I}}
\newcommand{\Groebner}{\mathbf{G}}

\newcommand{\tr}{^T}
\newcommand{\Gvec}{\mathbf{G}}
\newcommand{\fvec}{\mathbf{f}}
\newcommand{\kvec}{\mathbf{k}}

\newcommand{\mvec}{\mathbf{m}}
\newcommand{\nvec}{\mathbf{n}}
\newcommand{\svec}{\mathbf{s}}
\newcommand{\vvec}{\mathbf{v}}

\newcommand{\xvec}{\mathbf{x}}
\newcommand{\yvec}{\mathbf{y}}
\newcommand{\numvars}{v}
\newcommand{\numirreds}{\abs{\mvec}}
\newcommand{\numgens}{n}

\makeatletter
    \newcommand*\bigcdot{\mathpalette\bigcdot@{1}}
    \newcommand*\smtimes{\mathpalette\smtimes@{.7}}
    \newcommand*\bigcdot@[2]{\mathbin{\vcenter{\hbox{\scalebox{#2}{$\m@th#1\bullet$}}}}}
    \newcommand*\smtimes@[2]{\mathbin{\vcenter{\hbox{\scalebox{#2}{$\m@th#1\times$}}}}}
\makeatother

\newcommand{\spqr}{SP$\mathbb{Q}$R}
\DeclareMathOperator{\Elim}{Elim}

\title{Sampling Polynomial Rational Remainders with SP$\mathbb{Q}$R: A new Package for Polynomial Division and Elimination}

\author[a,b]{Vsevolod Chestnov,}
\author[c]{Giulio Crisanti}

\affiliation[a]{\bologna}
\affiliation[b]{\oxford}
\affiliation[c]{\edin}

\newcommand{\bologna}{
    Dipartimento di Fisica e Astronomia, Universit\`a di Bologna
    e INFN, Sezione di Bologna,
    via Irnerio 46, I-40126 Bologna, Italy.
}
\newcommand{\oxford}{
    Mathematical Institute, University of Oxford, OX2 6GG, United Kingdom
}
\newcommand{\edin}{Higgs Centre for Theoretical Physics, University of Edinburgh, James Clerk Maxwell Building,Peter Guthrie Tait Road, Edinburgh, EH9 3FD, United Kingdom}

\vspace{4cm}

\emailAdd{vsevolod.chestnov@maths.ox.ac.uk}
\emailAdd{g.crisanti@ed.ac.uk}

\abstract{
We introduce \spqr{}, a new \soft{Mathematica} package for the division and elimination of variables from polynomial systems. \spqr{} works by sampling and reconstructing results over finite fields, in an analogous manner to many state of the art Integration by Parts algorithms for Feynman integrals. This allows \spqr{} to effectively overcome expression swell during the construction of Gr\"obner bases, which in many cases is the major bottleneck in such computations.
Benchmarks on state of the art Macaulay resultants show that \spqr{} can deliver substantial gains over symbolic computer algebra workflows — reducing both runtime and memory footprint by multiple orders of magnitude. Likewise when applied to study Feynman integrals, we show how \spqr{} can be used to find previously unknown Landau singularities.
}

\allowdisplaybreaks
\begin{document}
\addtocontents{toc}{\protect\setcounter{tocdepth}{3}}

\maketitle

\input{sections/introduction}

\input{sections/theoretical_background}

\input{sections/program_description}
\input{sections/examples}

\input{sections/conclusions}

\subsection*{Acknowledgements}
We would like to thank
Giacomo Brunello,
Simon Caron-Huot,
Hjalte Frellesvig,
Mathieu Giroux,
Franz Herzog,
Pierre Lairez,
Luke Lippstreu,
Andrew McLeod,
Pierpaolo Mastrolia,
Sebastian Mizera,
Rafael Mohr,
Ben Page,
Tiziano Peraro,
Maria Polackova,
Sid Smith,
Bernd Sturmfels,
and
Felix Tellander
for many useful discussions and helpful comments on the manuscript.
We would like to especially thank Franz Herzog for generously providing substantial computational resources during the development of this project.

V.C.'s research is supported by the European Research Council (ERC) under the European Union's Horizon Europe research and innovation program grant agreement 101040760 (ERC Starting Grant \emph{FFHiggsTop}), and by the ERC Synergy Grant MaScAmp 101167287. 
G.C.'s research is supported by the United Kingdom Research and Innovation grant UKRI FLF MR/Y003829/1.
Views and opinions expressed are those of the authors only and do not necessarily reflect those of the European Union or the European Research Council. Neither the European Union nor the granting authority can be held responsible for them.
\newpage
\appendix

\bibliographystyle{JHEP}
\bibliography{biblio}
\end{document}

%% file: sections/introduction.tex
\section{Introduction}

Understanding and solving systems of polynomial equations is ubiquitous in mathematics, physics and beyond.
Out of the many algorithms one can employ when studying such problems, perhaps the most important is that of \emph{polynomial division}.
This algorithm forms the backbone of many of the most powerful tools known to process such systems.

Indeed, for systems of equations in multiple variables, one of the most common uses of polynomial division is the construction of Gr\"obner bases \cite{10.1145/1088216.1088219}.
With such a basis at hand many properties of a polynomial system become manifest: the number and dimensionality of its roots can be readily determined and variables can be systematically eliminated.
This last property is crucial for, among other things, a systematic algorithm to find the roots of polynomial systems.
Other important applications include computing syzygies, simplifying and solving multivariate algebraic constraints, as well as studying polynomial varieties systematically \cite{Cox:2015ode,CoxLittleOshea:2005,Sturmfels:2002}.

Beyond these general applications with vast scope, polynomial division and Gr\"obner basis algorithms enjoy many more specific applications.
In the field of scattering amplitudes alone these techniques have been applied to
integration by parts (IBPs) \cite{Gluza:2010ws,Schabinger:2011dz,Ita:2015tya,Larsen:2015ped,Agarwal:2020dye,Wu:2023upw} based on syzygy computations, as well as intersection theory inspired approaches~\cite{Page:2025gso};
exploration~\cite{Bitoun:2017nre} and implementation~\cite{Bertolini:2025zud} of parametric annihilators;
intersection number computations~\cite{Weinzierl:2020xyy,Fontana:2022yux,Fontana:2023amt,Brunello:2023rpq,Brunello:2024tqf};
spinor helicity computations \cite{DeLaurentis:2022otd,Campbell:2025ftx};
gravitational waveforms \cite{Brunello:2025cot};
Landau analysis algorithms \cite{Fevola:2023fzn,Helmer:2024wax};
integrand decompositions \cite{Zhang:2012ce,Mastrolia:2012an};
determination of annihilators of Feynman integrals~\cite{Muller-Stach:2011qkg, Lairez:2022zkj, delaCruz:2024xit} and their structures~\cite{Chestnov:2025whi};
as well as many other related problems \cite{Gasparotto:2023roh,Heller:2021qkz}.

Given their importance, considerable effort has been devoted to developing more efficient strategies for computing Gr\"obner bases \cite{FAUGERE199961,10.1145/780506.780516,FAUGERE1993329} and implementing these algorithms in fast and reliable computer codes \cite{DGPS,M2,msolve}.
Despite significant progress however, building Gr\"obner basis often remains a major bottleneck in practical applications.
Indeed in general their calculation is known to be demanding, with worst-case (saturated) upper bounds that grow doubly exponentially in the number of variables \cite{Hermann1926}.

Another important phenomenon that can make Gr\"obner basis computations challenging is \emph{expression swell}: during exact symbolic manipulations, intermediate results can grow by orders of magnitude, both in degree and in size, often far beyond those of the final output \cite{ARNOLD2003403,10.1007/3-540-51084-2_12}.
This process is greatly exacerbated when the roots of the given polynomial systems depend on many symbolic parameters.
In practice, uncontrolled swell can render otherwise modest reductions intractable.

Aside from polynomial algebra, expression swell is also frequently a problem in Integration by Parts algorithms (IBPs) for Feynman integrals \cite{Tkachov:1981wb,Chetyrkin:1981qh,Laporta:2001dd}.
Essentially IBPs amount to performing Gau\ss{}ian elimination on very large systems of equations, where physical parameters often cause intermediate expressions to become unmanageable.
In the last decade, this problem has been mitigated to great success with the introduction of \emph{finite field sampling and reconstruction} methods \cite{Kauers:2008zz,Kant:2013vta,vonManteuffel:2014ixa,Peraro:2016wsq,Peraro:2019svx,Smirnov:2019qkx,Klappert:2019emp,10.1145/3717582.3717588}. 

The key idea behind these approaches is to substitute all parameters for integers and perform all computations numerically, modulo a (large) prime number.
By performing the same computation on multiple numerical slices, the full parametric dependence of the output can be restored via interpolation methods.
Any remaining numerical coefficients can then be lifted back to the rational numbers via Wang's algorithm \cite{Wang:1981} and the Chinese remainder theorem if necessary \cite{Peraro:2016wsq,Peraro:2019svx}.

This approach presents three important advantages.
Firstly, the modulo arithmetic ensures that all integer expressions are capped in size.
This avoids expression swell even on the numerical slices. Secondly, any complicated cancellations in the algorithm's output happen numerically, and before any interpolation/reconstruction.
This in many situations effectively avoids the need for heavy symbolic processing steps which are required to see algebraic simplifications in many implementations.
Finally this strategy is massively parallelisable, as numerical evaluations of sample points are fully independent calculations.
This allows for effective scaling against available computer resources.

In this work we concretely demonstrate that finite field sampling and reconstruction methods can be extremely effective in tackling heavily parametric
problems in polynomial algebra.
Indeed, the construction of Gr\"obner bases can be recast as solving linear systems through the construction of suitably large Macaulay systems \cite{947c5834c2744a6f9ab0a3a3d98eef20,Buchberger2017,cryptoeprint:2025/793}.
Specifically one multiplies the generators of the ideal by a finite set of monomials, assembles the resulting relations into a matrix, and performs Gau\ss{}ian elimination to read off the remainders of the target polynomials.
By performing these operations with finite field sampling and reconstruction algorithms, it is thus possible to bypass the intermediate expression swells that many algebraic Gr\"obner basis algorithms suffer from \cite{ARNOLD2003403,10.1007/3-540-51084-2_12}.

The extra structure present in (zero-dimensional) polynomial ideals also allows for the introduction and extensive usage of \emph{companion} (often also denoted \emph{multiplication}) \emph{matrices} \cite{Sturmfels:2002,CoxLittleOshea:2005,Cox:2015ode,Huang:2015yka,Jiang:2017phk,Telen:2020thesis,FebresCordero:2023pww,Brunello:2024tqf}.
These matrices provide an elegant method to polynomially reduce any multivariate \emph{rational} function, beyond just polynomials.
Furthermore, they can be used to eliminate variables from polynomial systems by analysing their respective eigenvalue systems \cite{alma9911250933502466,Cox:2020}.
By their nature, operations with and on companion matrices can also be ported to a finite fields pipeline, allowing for a clean integration with any other algorithms.

In this paper we present the \soft{Mathematica} package \spqr{}, or Sampling Polynomial Rational Remainders in full.
\spqr{}'s main goal is to perform polynomial division and variable elimination in polynomial systems using a fully finite fields workflow, designed to never encounter intermediate expression swells.  
 
As its backbone \spqr{} utilises extensively the package \soft{FiniteFlow} \cite{Peraro:2019svx}, which supports many basic operations on functions, lists and matrices over finite fields.
\spqr{} then repackages these low level operations into high level user accessible commands aimed at the analysis of polynomial systems.
Only the final desired output is numerically sampled and reconstructed, ensuring that any complicated cancellations always happen numerically.
Crucially this also means that \spqr{} never symbolically builds an explicit Gr\"obner basis, as this is nearly always an intermediate step in most computations.
 
Like all algorithms, there are certain classes of problems where \spqr{}'s approach is best suited.
Crudely speaking it is useful to distinguish between the ``variable complexity" and ``parameter complexity" of polynomial systems: variable complexity is set by the number of reduction variables and the degrees in which they appear, solutions then depend on the remaining symbolic parameters.
Since intermediate expression swell is driven primarily by parametric coefficients, \spqr{} is most effective on systems with rich parameter dependence and moderate variable complexity. 

We argue this regime is common in high energy scattering amplitudes, where one is often interested in polynomial systems with multiple kinematic scales, which generate complicated parameter dependence.
In such cases we find that \spqr{} can be a very effective tool to analyse these systems, with multiple order of magnitude improvements in both computation time and RAM usage compared to publicly available computer algebra codes \cite{DGPS,M2,msolve}.
We also find similar improvements when considering the construction of Macaulay resultants, which by nature are also heavily parameter dependent. 

This work is structured as follows: In \cref{sec:theoretical_background} a (gentle) introduction to all the necessary theoretical background pertaining to \spqr{} is presented.
We begin with a review of univariate and multivariate polynomial division, before discussing Gr\"obner bases, the Gau\ss{}ian elimination approach, as well as the companion matrix formalism.
Finally, we show how all these concepts can be ported to a finite fields reconstruction setting.
Section \ref{sec:prog_install} focuses on using \spqr{}, including installation instructions and a quick start guide for many important workflows.
In \cref{sec:examples} we apply \spqr{} to some more difficult problems.
We first consider the computation of state of the art Macaulay resultants, and provide benchmarks against other computer algebra systems. In these tests we find at least 5-6 orders of magnitude improvements in compute time and 3-4 orders in memory usage.
We then consider the study of Feynman integrals, where we show \spqr{} can be used to find previously unknown Landau singularities.
Finally, concluding remarks as well as a future outlook for \spqr{}'s development is provided in \cref{sec:conclusions}.

%% file: sections/theoretical_background.tex
\newpage
\section{Theoretical Background}\label{sec:theoretical_background}
As already anticipated, the core functionality behind \spqr{} revolves around understanding the structure and eliminating variables from systems of equations. This is primarily achieved through the use of polynomial division as well as related operations, which are performed over finite fields in \spqr. To this end, in this section we review the theoretical background underlying \spqr's functionality, namely: polynomial division, Gau\ss{}ian elimination, companion matrices and elimination theory. Finally, how all these operations can be passed through finite field reconstruction algorithms is discussed.
\subsection{Review of Polynomial Division}
\subsubsection{Univariate Polynomial Division}\label{sec:univariate_poly_div}
Given two univariate polynomials $f(x)$ and $p(x)$, polynomial division is grounded in the decomposition
\begin{equation}\label{eq:univ_decomp}
    f(x) = q(x)\, p(x) + r(x)\,,
\end{equation}
where crucially $\deg(r)<\deg(p)$.
The polynomials $q(x)$ and $r(x)$ are respectively known as the quotient and remainder.
The fact that the form in~\cref{eq:univ_decomp} can always be reached is easiest shown by direct construction — one proceeds by rearranging the equation for $p(x)$ such that its leading monomial is isolated, and then repeatedly substitutes this equation into $f(x)$ as much as possible.
As an example, if
\begin{align}\label{eq:univ_example}
    f\brk{x} = x^3 + a x^2 - \brk{4 + 2 a} x + 1
    \>,
    \quad
    p\brk{x} = x^2 - 2 x - 1
    \>,
\end{align}
The second equation can be rearranged as $x^2 = p(x) + 2x+1$, and any term of degree $>2$ in $f(x)$ can be substituted as
\begin{equation}\label{eq:univ_decomp_explicit}
    \begin{aligned}
    f(x) &= x\, (p(x) + 2x + 1) + a\, (p(x)+2x+1) - (4 + 2a)x + 1\,, \\
    &=p(x)(x+a) + 2x^2 + a -3x + 1\,, \\
    &=p(x)(x+a) + 2(p(x)+2x+1) + a -3x + 1\,,\\
    &=p(x)(x+a+2) + x+a+3\,.
    \end{aligned}
\end{equation}
Thus, for this specific example $q(x) = x+a+2$ and $r(x)=x+a+3$.
It is straightforward to see that \textit{any} polynomial $f(x)$, using the above strategy will result in a remainder term of degree $<2$, and thus will be a linear combination of the two \textit{irreducible} monomials\footnote{
    Another term for the set of irreducible monomials in the literature is the \textit{staircase} of an ideal $\Ideal$, owning to the shape that the exponents of irreducible monomials \brk{in the multivariate case} fill out in the integer lattice of all possible monomial exponents.
}
\begin{align}
    \mons = \brc{m_1, \> m_2} = \brc{x, \> 1}\,,
    \qquad r(x) = m_1 + (a+3)\,m_2\,.
\end{align}
Indeed for a generic univariate polynomial $p\brk{x}$ of degree $\deg(p(x)) = d$, there will always be $\numirreds = d$ irreducible monomials given by $\{x^{d-1},\ldots,x^0\}$. 

In many applications, only the remainder of the polynomial division is of interest.
For this reason, throughout this work we will focus on the this term, which can be expressed in modulo notation as
\begin{equation}
    f(x) = r(x) \mod p(x).
\end{equation}

\subsubsection{Univariate Rational Function Division}
Polynomial division can also be extended beyond simply polynomials to rational functions $g(x)/f(x)$.
To do this, it is necessary to define the polynomial inverse of $f\brk{x}$. This is a new polynomial $f_\mathrm{inv}\brk{x}$ such that\footnote{The inverse is well defined when $\gcd(f,p)=1$, i.e. when $f$ and $p$ do not have a common root.}
\begin{equation}\label{eq:univ_inverse_def}
    f_\text{inv}(x) \, f(x) = 1 \mod p(x)\,.
\end{equation}
Determination of $f_\text{inv}(x)$ is algorithmic and can be computed using the extended Euclidean algorithm, via ansatz, or instead via companion matrices, which is the approach implemented in \spqr{} discussed later in \cref{sec:companion_matrices}.
For $f(x)$ above the inverse is given by
\begin{equation}\label{eq:univ_inverse_explicit}
    f_\text{inv}(x) = \frac{a+5}{a^2+8 a+14}-\frac{x}{a^2+8 a+14}\,,
\end{equation}
which can be verified by direct substitution.
We note that, while the parameter $a$ is allowed to appear in the denominator, the variable $x$, which the division is performed against, only appears in the numerator, thus rendering~\cref{eq:univ_inverse_explicit} a polynomial in it\footnote{
    In other words, the remainder always belongs to the ring of polynomials in $x$ with rational function coefficients in $a$, see more in~\cref{sec:which_ring}.
}.

With the inverse defined, it is straightforward to tackle the division of rational functions.
Indeed, given any $g(x)/f(x)$ one can compute
\begin{equation}
    \frac{g(x)}{f(x)}=f_\text{inv}(x)g(x) \mod p(x)\,,
    \label{eq:univ_rat_reduction}
\end{equation}
and perform further polynomial reductions on the right hand side if necessary.
\subsubsection{Univariate polynomial division as linear system solving}\label{sec:univ_row_reduction}
It is also possible to recast the problem of finding polynomial remainders as solving a linear system of equations.
Given the polynomials in~\cref{eq:univ_example}, we can generate a system of identities
\begin{equation}
    \left.
\begin{aligned}
    x^2 - 2 x - 1 &= 0
    \>
    \\
    x^3 - 2 x^2 - x &= 0
    \>
    \\
     &\vdots
\end{aligned}
    \> \right\} \hspace{-8pt} \mod p(x)\,,
\end{equation}
built by considering $x^n\,p(x)=0 \mod p(x)$.
We add to this list the defining equation for the polynomial
$f\brk{x}$ as
\begin{align}
    f\brk{x} - x^3 - a x^2 + \brk{4 + 2 a} x - 1 = 0
    \>,
\end{align}
and organize the whole system in matrix form, called the \textit{Macaulay matrix}~\cite{Macaulay:1916}:
\begin{equation}\label{eq:1var_macaulay_matrix}
    \NiceMatrixOptions{code-for-first-col = \scriptstyle, code-for-last-col = \scriptstyle}
    \begin{bNiceMatrix}[margin, first-col]
        f \rightarrow
        & 1 & -1 & -a & 4 + 2 a & -1
        \\
        & \mzero & 1 & -2 & -1 & \mzero
        \\
        & \mzero & \mzero & 1 & -2 & -1
        \CodeAfter
            \tikz \draw[
                line width=.4pt, gr,
            ]
                (1|-2) -- (6|-2)
                (1-|2) -- (4-|2)
            ;
    \end{bNiceMatrix}
    \cdot
    \begin{bNiceMatrix}
        f \\
        x^3 \\
        x^2 \\
        x \\
        1
        \CodeAfter
            \tikz \draw[
                line width=.4pt, gr,
            ]
                (1|-2) -- (2)
            ;
    \end{bNiceMatrix}
    = 0 \mod p(x)
    \>.
\end{equation}
To solve this system we may perform Gau\ss{}ian elimination to bring the matrix to reduced row echelon form, which reads
\begin{equation}
    \NiceMatrixOptions{code-for-first-col = \scriptstyle, code-for-last-col = \scriptstyle}
    \begin{bNiceMatrix}[margin, first-col]
        \CodeBefore
        \Body
            & 1 & \mzero & \mzero & -1 & -3 - a
            \\
            & \mzero & 1 & \mzero & -5 & -2
            \\
            p \rightarrow & \mzero & \mzero & 1 & -2 & -1
        \CodeAfter
            \tikz \draw[
                line width=.4pt, gr,
            ]
                (1|-2) -- (6|-2)
                (4|-1) -- (4|-6)
                (1-|2) -- (4-|2)
            ;
    \end{bNiceMatrix}
    \cdot
    \begin{bNiceMatrix}[last-col]
        \CodeBefore
        \Body
            f & \\
            x^3 & \\
            x^2 & \\
            x & \leftarrow m_1 \\
            1 & \leftarrow m_2
        \CodeAfter
            \tikz \draw[
                line width=.4pt, gr,
            ]
                (1|-2) -- (2)
                (1|-4) -- (2|-4)
            ;
    \end{bNiceMatrix}
    = 0 \mod p(x)
    \>,
\end{equation}
where we have highlighted the irreducible monomials $\brc{m_1, m_2} = \brc{x, 1}$ and the corresponding blocks of the system with their coefficients.
The result of the polynomial division can now be read off the top row as $f(x) - x -3-a=0\mod p(x)$, and thus once again we find $r(x) = x + a +3$. 

This method is no different to the algorithm presented in \cref{eq:univ_decomp_explicit}. Nevertheless, it provides a clean formulation of the problem from a computational perspective and forms the basis for \spqr{}'s approach to polynomial division.
\subsubsection{Vanishing sets and polynomial remainders}\label{sec:vanishing_sets_univariate}
There exists an important link between the roots of polynomial systems and the remainders of polynomial division.
Given a polynomial $p(x)$, the \textit{vanishing set} $V(p)$ is a subset of the complex plane $\Complex$ defined as
\begin{equation}
    V(p) = \bigbrc{x \in \Complex \,\big|\, p(x)=0}\,,
\end{equation}
namely as the set of roots of $p$.
Suppose that $x^* \in V(p)$ is such a root.
From~\cref{eq:univ_decomp} it is simple to see that the value of the function $f$ at this point concides with the value of the remainder $r$, namely
\begin{equation}
    f(x^*) = r(x^*)\,.
\end{equation}
Thus, identities that are true modulo $p(x)$ become exact when considering points in $V(p)$:
\begin{equation}\label{eq:vanishing_sets_univar}
    f(x)=r(x)\mod p(x) \Longrightarrow f(x^*) =r(x^*) \quad \forall \,\, x^* \in V(p)\,.
\end{equation}
This relation allows one to relate much of the technology developed for building polynomial remainders, to understanding the solution structure of polynomial systems of equations.
Indeed this relation (and its multivariate counterpart given in~\cref{eq:vanishing_sets_multivar}) are what allow for \spqr's division algorithms to be powerful tools when studying systems of polynomial equations.

As a simple example of~\cref{eq:vanishing_sets_univar} in action, we can consider $f(x)$ and $p(x)$ given in~\cref{eq:univ_example}.
We have
\begin{equation}
    V(p) = 1 \pm \sqrt{2}\,,
\end{equation}
which can be plugged into $f$ to obtain
\begin{equation}\label{eq:f_vanishing_explicit}
    f(V(p)) = -\left(1\pm \sqrt{2}\right) (2 a+4)+\left(1\pm \sqrt{2}\right)^2 a+\left(1\pm\sqrt{2}\right)^3+1 = 4 + a \pm \sqrt2\,.
\end{equation}
It is important to note that the last equality required an algebraic simplification step to expand and cancel various roots.
The same result can be reached avoiding this latter step by substituting $V(p)$ directly into $r(x) = 3 + a + x$,%
\begin{equation}
    r(V(p)) = 4 + a \pm \sqrt{2}\,.
\end{equation}
This relationship also holds for inverses and rational functions: from~\cref{eq:f_vanishing_explicit} we have
\begin{equation}
    \frac{1}{f(V(p))}=\frac{1}{4+a\pm \sqrt{2}} = \frac{4+a\mp \sqrt{2}}{14+8a+a^2}\,,
\end{equation}
which can likewise straightforwardly be obtained by considering $f_\text{inv}(V(p))$ taken from~\cref{eq:univ_inverse_explicit}.
\subsubsection{Multivariate Polynomial Division}
In most cases of practical interest, one has to deal with polynomial systems $\{p_1(\xvec),\ldots,p_\numgens(\xvec)\}$ in multiple variables $\xvec=\{x_1,\ldots,x_\numvars\}$.
Suppose the goal is to perform polynomial division on a function $f(\xvec)$.
In principle we can write the decomposition
\begin{equation}\label{eq:poly_div_multivariate_naive}
    f(\xvec) = \sum_{j=1}^{\numgens} p_j(\xvec)\, q_j(\xvec) + r(\xvec)\,.
\end{equation}
In direct analogy to \cref{eq:univ_decomp}. The polynomials $\{ p_1, \ldots, p_\numgens \}$ can be viewed as generators of an \textit{ideal} $\Ideal = \langle p_1, \ldots, p_\numgens \rangle$, which is defined such that
\begin{equation}
    \sum_{j=1}^{\numgens}p_j(\xvec)\, q_j(\xvec) \in \Ideal \quad \forall \,\, q_j(\xvec)\,.
\end{equation}
Focusing once again on the remainder, \cref{eq:poly_div_multivariate_naive} can thus be rewritten as
\begin{equation}
    f(\xvec) = r(\xvec) \mod \Ideal \,.
\end{equation}
It is also straightforward to see that
\begin{equation}
    f(\xvec) = 0 \mod \Ideal \Longleftrightarrow f(\xvec) \in \Ideal\,.
\end{equation}
Indeed the relationships between the roots of polynomial systems and polynomial remainders, discussed in~\cref{sec:vanishing_sets_univariate}, generalise straightforwardly to the multivariate case.
If $\xvec^* \in V(\Ideal) = V(p_1,\ldots,p_\numgens)$, where
\begin{align}
    V\brk{p_1, \ldots, p_{\numgens}}
    =
    \bigbrc{
        \xvec \in \Complex^\numvars \,\big|\,
        p_1\brk{\xvec} = \ldots = p_{\numgens}\brk{\xvec} = 0
    }
    \>,
\end{align}
then it is clear again that
\begin{equation}\label{eq:vanishing_sets_multivar}
    f(\xvec) = r(\xvec) \mod \Ideal \Longrightarrow f(\xvec^*) = r(\xvec^*) \quad \forall\,\,\mathbf{x^*}\in V(\Ideal)\,.
\end{equation}
If $V(\Ideal)$ is composed of isolated points, then $\Ideal$ is said to be \emph{zero-dimensional}. If higher dimensional solutions are present then instead the ideal has positive dimension.

Despite many similarities, multivariate polynomial division presents significantly more challenges and ambiguities compared to the univariate setting which need to be addressed for the procedure to be useful.

For polynomials in one variable, there is the implicit assumption that a monomial $x^j$ is considered ``worse" or ``higher weight" than $x^i$ if $j>i$.
Indeed, the univariate division algorithm can be seen as a substitution of monomials in an attempt to reduce the weight of $f(x)$ as much as possible, with the \emph{monomial ordering} $x^i>x^{i-1}>\ldots>x>1$.
For more than one variable, there is no unique canonical choice for the weight ordering, and many exist.
In two variables $\xvec=\{x,y\}$, if we assume $x>y$, one common choice is lexicographic (dictionary) ordering, which is given by
\begin{equation}\label{eq:lexicographic_order}
\ldots >xy^\infty>\ldots >xy>x> y^\infty \ldots >y>1\,,
\end{equation}
where by $\infty$ we mean an arbitrarily large monomial power.
Another common choice is degree reverse lexicographic, where the sum of the powers primarily determines the weight, and among monomials of equal total degree, exponents are compared from the last variable backward, with the monomial having the \emph{smaller} exponent in the first differing variable considered larger:
\begin{align}
    \ldots > y^3 > x^2 > x y > y^2 > x > y > 1\,.
\end{align}
Further frequent choices include degree lexicographic and various types of elimination orderings.
In computer implementations, it is convenient to represent monomial orders by means of a \textit{weight matrix} $W$, which acts on exponent vectors to produce numerical weights used for comparison.
Given two monomials\footnote{
    Here and in the following we use the multi-index notation $\xvec^\nvec = x_1^{n_1} x_2^{n_2} \ldots x_{\numvars}^{n_{\numvars}}$.
} $\xvec^\kvec$ and $\xvec^\nvec$ one compares the entires of the corresponding weight vectors $W \cdot \kvec$ and $W \cdot \nvec$ component-wise, starting from the first entry and proceeding to the next only if the comparison so far has not resolved the ordering..
For illustration, in case of five variables the weight matrices for the lexicographic, degree lexicographic, and degree reverse lexicographic take the following shape:
\begin{equation}
    \brc{W_\mathrm{lex}, W_\mathrm{deglex}, W_\mathrm{degrevlex}}
    =
    \Bigbrc{
        \begin{bNiceMatrix}
            1 & \mzero & \mzero & \mzero & \mzero
            \\
            \mzero & 1 & \mzero & \mzero & \mzero
            \\
            \mzero & \mzero & 1 & \mzero & \mzero
            \\
            \mzero & \mzero & \mzero & 1 & \mzero
            \\
            \mzero & \mzero & \mzero & \mzero & 1
        \end{bNiceMatrix}
        \>,
        \quad
        \begin{bNiceMatrix}
            1 & 1 & 1 & 1 & 1
            \\
            1 & \mzero & \mzero & \mzero & \mzero
            \\
            \mzero & 1 & \mzero & \mzero & \mzero
            \\
            \mzero & \mzero & 1 & \mzero & \mzero
            \\
            \mzero & \mzero & \mzero & 1 & \mzero
        \end{bNiceMatrix}
        \>,
        \quad
        \begin{bNiceMatrix}
            1 & 1 & 1 & 1 & 1
            \\
            \mzero & \mzero & \mzero & \mzero & -1
            \\
            \mzero & \mzero & \mzero & -1 & \mzero
            \\
            \mzero & \mzero & -1 & \mzero & \mzero
            \\
            \mzero & -1 & \mzero & \mzero & \mzero
        \end{bNiceMatrix}
    }
    \>.
    \label{eq:weight_matrices}
\end{equation}
To see these orderings in action, consider the following three monomials:
\begin{equation}
    \renewcommand{\arraystretch}{1.2}
    \begin{NiceArray}{lW{c}{2.7cm}W{c}{2.7cm}c}
        & \text{monomial} & \text{exponent vector} & \text{degree}
        \\
        \hline
        p_1
        & x_1 x_2 x_3 x_4^2 x_5^3
        & \brc{1, 1, 1, 2, 3}
        & 8
        \\
        p_2
        & x_1 x_2 x_3 x_4^3 x_5
        & \brc{1, 1, 1, 3, 1}
        & 7
        \\
        p_3
        & x_1 x_2 x_3^2 x_4 x_5^2
        & \brc{1, 1, 2, 1, 2}
        & 7
    \end{NiceArray}
\end{equation}
Using the three weight matrices from~\cref{eq:weight_matrices}, these mononmials are ordered in three different ways:
\begin{equation}
    \begin{NiceArray}{rW{c}{2.5cm}}
        \mathrm{lex}
        & p_3 > p_2 > p_1
        \\
        \mathrm{deglex}
        & p_1 > p_3 > p_2
        \\
        \mathrm{degrevlex} 
        & p_1 > p_2 > p_3
    \end{NiceArray}
\end{equation}

However, even with a well defined choice of monomial ordering, polynomial division still suffers from an important problem: the remainder is not defined uniquely. This property can already be seen in very simple examples. Let
\begin{equation}\label{eq:ideal_2var_explicit}
    \Ideal = \langle xy-x,xy-y-1\rangle\,,
\end{equation}
be a 2-variate ideal and suppose the goal is to compute $f(x,y)=x\,y \mod \Ideal$ in lexicographic ordering.

By applying the first equation in $\Ideal$ one would conclude that $f(x,y)=x \mod \Ideal$, as $xy>x$, and no further substitutions are possible.
However, if instead one used the second entry in $\Ideal$, then likewise $x y > y+1$, and thus one would instead reach the different result $f(x,y) = y + 1 \mod \Ideal$.
Thus, a na\"ive approach to polynomial division in the multivariate setting, although correct, is of little practical use.
\subsubsection{Gr\"obner Bases}
\begin{figure}[H]
    \centering
    \includegraphicsbox[width=0.3\textwidth]{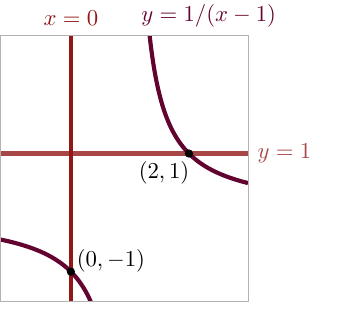}
    \hspace{1cm}
    \includegraphicsbox[width=0.3\textwidth]{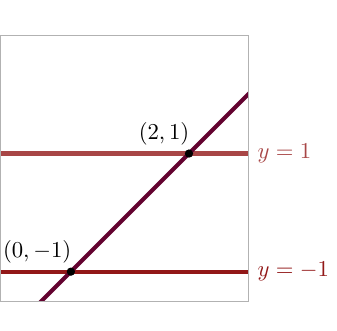}
    \caption{
        The left panel shows the vanishing set of the ideal~\cref{eq:ideal_2var_explicit}, while the right one depicts that of its Gr\"obner basis~\cref{eq:multivar_example_gb}; the two sets coincide.
    }
    \label{fig:vanishing_set}
\end{figure}

The uniqueness of polynomial divisions in the multivariate setting can be restored by introducing Gr\"obner bases \cite{10.1145/1088216.1088219}.
A Gr\"obner basis $\Groebner$ is a special generating set of an ideal $\Ideal$ with many desirable characteristics.
For the purposes of this work, the most important property is that polynomial division using the new generators $\Groebner$ is no longer ambiguous (for a given monomial order). Furthermore, as with any set of generators, $\Gvec$ satisfies $V(\Groebner)=V(\Ideal)$ \cite{10.1145/1088216.1088219,CoxLittleOshea:2005,Cox:2015ode,Sturmfels:2002}.
For example, taking $\Ideal$ from~\cref{eq:ideal_2var_explicit}, its respective Gr\"obner basis (in lexicographic order) is given by
\begin{equation}
    \Groebner = \{ g_1, g_2 \} = \{ y^2 -1 ,x-y-1 \}\,, \qquad \Ideal.
    \label{eq:multivar_example_gb}
\end{equation}
It is straightforward to verify that
\begin{equation}\label{eq:ideal_explicit_roots}
    V(\Ideal) = V(\Groebner) = \{x=0\,,\,y=-1\} \,\cup\, \{x=2\,,\,y=1\}\,.
\end{equation}
as can also be seen visually in \cref{fig:vanishing_set}. Indeed one can also write
\begin{equation}
    \Ideal = \langle \Groebner \rangle\,.
\end{equation}
Polynomial division of $f(x,y)$ using $\Groebner$ now results in the unambiguous result
\begin{equation}\label{eq:gb_pdiv}
    f(x,y) = y+1 \mod \Ideal\,,
\end{equation}
which can be verified by substituting the second followed by the first equation of $\Groebner$ into $f$.
For more complicated examples, a Gr\"obner basis ensures that any ``reduction order" will result in the same unique answer.
For example, if $f(x,y) = x\, y^2$, then
\begin{equation}
    \left.
\begin{aligned}
        f(x,y) &= x = y + 1 \\
        f(x,y) &= (y+1)\,y^2 = y+1
\end{aligned}
    \quad \right\} \hspace{-8pt} \mod \Ideal\,,
\end{equation}
where different substitution orders have been used in the two identities. With a Gr\"obner basis at hand it is also possible to straightforwardly determine the irreducible monomials of $\Ideal$. In particular, an infinite number of irreducible monomials can be show to imply that $\Ideal$ is not zero-dimensional. This can be a useful test to determine the nature of the roots of a polynomial system.

There exist currently multiple state of the art algorithms \cite{10.1145/1088216.1088219,FAUGERE199961,FAUGERE1993329,10.1145/780506.780516} and computer algebra implementations \cite{DGPS,M2,msolve} of Gr\"obner basis computations.
Nevertheless their computation in many cases can be challenging, and (saturated) upper bounds on complexity are known to scale very poorly \cite{Hermann1926}.
\subsection{Multivariate Polynomial Division as Linear System Solving}\label{sec:multi_row_reduction}
Beyond the standard Gr\"obner basis computation algorithms, there exist other methods to compute polynomial remainders uniquely.
Remarkably, casting the problem as a linear system, discussed in~\cref{sec:univ_row_reduction} generalises straightforwardly to the multivariate setting.

To this end let us take $\Ideal~=~\vev{x y - x, x y - y - 1}$ as in \cref{eq:ideal_2var_explicit}, and consider the reduction of $f\brk{x, y} = x\, y^2$ with respect to $\Ideal$, in lexicographic ordering.
We may construct a system of linear equations by considering
\begin{equation}
    x^i y^j\, \Ideal = 0 \mod \Ideal\,,
\end{equation}
as well as the defining equation for $f$
\begin{equation}
    f(x,y) - x y^2 = 0\,.
\end{equation}
All the above identities can once again be neatly packaged into a Macaulay matrix, given by
\begin{equation}\label{eq:2var_macaulay_matrix}
    \NiceMatrixOptions{code-for-first-col = \scriptstyle, code-for-last-col = \scriptstyle}
    \begin{bNiceMatrix}[margin, first-col]
        f \rightarrow
        & 1 & \mzero & \mzero & -1 & \mzero & \mzero & \mzero & \mzero & \mzero
        \\
        & \mzero & 1 & \mzero & \mzero & -1 & -1 & \mzero & \mzero & \mzero
        \\
        & \mzero & \mzero & \mzero & 1 & \mzero & \mzero & -1 & -1 & \mzero
        \\
        & \mzero & \mzero & \mzero & \mzero & 1 & \mzero & \mzero & -1 & -1
        \\
        & \mzero & 1 & -1 & \mzero & \mzero & \mzero & \mzero & \mzero & \mzero
        \\
        & \mzero & \mzero & \mzero & 1 & -1 & \mzero & \mzero & \mzero & \mzero
        \\
        & \mzero & \mzero & \mzero & \mzero & 1 & -1 & \mzero & \mzero & \mzero
        \CodeAfter
            \tikz \draw[
                line width=.4pt, gr,
            ]
                (1|-2) -- (10|-2)
                (1-|2) -- (9-|2)
            ;
    \end{bNiceMatrix}
    \cdot
    \begin{bNiceMatrix}
        f \\
        x^2 y \\
        x^2 \\
        x y^2 \\
        x y \\
        x \\
        y^2 \\
        y \\
        1
        \CodeAfter
            \tikz \draw[
                line width=.4pt, gr,
            ]
                (1|-2) -- (2)
            ;
    \end{bNiceMatrix}
    =
    0 \mod \Ideal
    \>,
\end{equation}
where we have truncated the system at the monomial powers $i + j \leq 1$ \brk{see~\cref{sec:further_details} for further discussion of this truncation strategy}.
We note that the chosen ordering dictates the positions of the monomials in the vector, and thus the columns in the linear system.

Bringing this system to row-reduced echelon form unravels the algebraic structure of the ideal: from the first row one can immediately read that $f(x,y) - y- 1 =0 \mod \Ideal$, in agreement with~\cref{eq:gb_pdiv}.
Furthermore, one can identify the two irreducible monomials $\brc{m_1, m_2} = \{y,1\}$ corresponding to the last two columns, as well as the two generators of the Gr\"obner basis~\cref{eq:multivar_example_gb}, appearing in the last two rows:
\begin{equation}\label{eq:macaulay_2var_row_reduced}
    \NiceMatrixOptions{code-for-first-col = \scriptstyle, code-for-last-col = \scriptstyle}
    \begin{bNiceMatrix}[margin, first-col]
        \CodeBefore
        \Body
            & 1 & \mzero & \mzero & \mzero & \mzero & \mzero & \mzero & -1 & -1
            \\
            & \mzero & 1 & \mzero & \mzero & \mzero & \mzero & \mzero & -2 & -2
            \\
            & \mzero & \mzero & 1 & \mzero & \mzero & \mzero & \mzero & -2 & -2
            \\
            & \mzero & \mzero & \mzero & 1 & \mzero & \mzero & \mzero & -1 & -1
            \\
            & \mzero & \mzero & \mzero & \mzero & 1 & \mzero & \mzero & -1 & -1
            \\
            g_2 \rightarrow & \mzero & \mzero & \mzero & \mzero & \mzero & 1 & \mzero & -1 & -1
            \\
            g_1 \rightarrow & \mzero & \mzero & \mzero & \mzero & \mzero & \mzero & 1 & \mzero & -1
        \CodeAfter
            \tikz \draw[
                line width=.4pt, gr,
            ]
                (1|-2) -- (10|-2)
                (8|-1) -- (8|-10)
                (1-|2) -- (8-|2)
            ;
            \UnderBrace[yshift=3pt]{7-8}{7-9}{\substack{\text{decomposition} \\ \text{coefficients}}}
    \end{bNiceMatrix}
    \cdot
    \begin{bNiceMatrix}[last-col]
        \CodeBefore
        \Body
        f & \\
        x^2 y & \\
        x^2 & \\
        x y^2 & \\
        x y & \\
        x & \\
        y^2 & \\
        y & \leftarrow m_1 \\
        1 & \leftarrow m_2
        \CodeAfter
            \tikz \draw[
                line width=.4pt, gr,
            ]
                (1|-2) -- (2)
                (1|-8) -- (2|-8)
            ;
    \end{bNiceMatrix}
    =
    0 \mod \Ideal \,.
\end{equation}

It is important to note that this method, whilst simple in nature, does not escape the complexity constraints that ordinary Gr\"obner basis computations face.
Unlike the one variable case, the number of equations which one needs to add to the Macaulay matrix to obtain a full reduction is unclear a priori.
Upper (saturated) bounds on the system size are known, and similarly to direct Gr\"obner basis algorithms they scale poorly \cite{Hermann1926,Buchberger2017}.
Because of these limitations, this algorithm, whilst already known, has received restricted interest due to its perceived inefficiency in general settings \cite{Buchberger2017}.
Nevertheless, a variant of this approach is used extensively in the state of the art algorithms~\cite{FAUGERE199961}.

The core innovation in \spqr's implementation is that in many cases of interest the Gau\ss{}ian elimination strategy can be very competitive with symbolic approaches in computational polynomial algebra.
This is due to the importation of novel state of the art algorithms and computer implementations for Gau\ss{}ian elimination from the scattering amplitudes community, which are discussed in detail in~\cref{sec:theoretical_implementation}.

Indeed, there are many parallels between Integration by Parts (IBP) techniques for Feynman integrals \cite{Tkachov:1981wb,Chetyrkin:1981qh,Laporta:2001dd} and multivariate polynomial division: the Macaulay system can be thought of as equivalent to an \textit{IBP system}, likewise irreducible monomials are analogous to \textit{master integrals} and the operation of polynomial reduction is equivalent to \textit{IBP reduction}.
\subsection{Companion Matrices}\label{sec:companion_matrices}
In the example considered in~\cref{sec:multi_row_reduction}, a Macaulay system up to weight $i + j \leq 1$ was generated to successfully reduce $f(x,y) = x\, y^2$.
Clearly, a much larger Macaulay system would have been required for the reduction of another polynomial with higher powers, say $f(x,y) = x^{13} y^{27}$.
This, in turn, would have resulted in a more computationally expensive row reduction step in order to achieve a successful polynomial reduction.

The companion matrix formalism solves this problem for zero-dimensional ideals by providing a direct way to recover the remainders of \textit{any} polynomial division without the need to generate unnecessarily large Macaulay systems.

\subsubsection{Basics of the Formalism}
Given a polynomial ideal $\Ideal$ (and a monomial ordering), let us assume a basis of irreducible monomials
\begin{align}
    \mvec = \brc{m_1, \ldots, m_{\numirreds}}
\end{align}
of length $\numirreds$\footnote{Here it is assumed that the $\Ideal$ is zero-dimensional, which ensures that $\mvec$ is of finite length.}.
For any polynomial $f(\xvec)$, its respective companion matrix can be built as follows: one multiplies $f(\xvec)$ with each irreducible monomial $m_i$ and performs polynomial division modulo $\Ideal$. The resulting remainders will be once again a linear combination of the irreducible monomials $\mvec$. This information can be neatly packaged into a matrix equation as
\begin{align}
    m_i \> f\brk{\xvec} = \sum_{j} \brk{M_{f(\xvec)}}_{ij} \> m_j \mod \Ideal
    \>.
    \label{eq:cmat_definition}
\end{align}
The $\numirreds \times \numirreds$ matrix $M_{f(\xvec)}$ is known as the \emph{companion}/\emph{multiplication matrix} associated with multiplication by $f\brk{\xvec}$.\footnote{Companion matrices can be thought of as analogous to differential equation matrices when considering IBP systems for Feynman integrals.}
In other words, the companion matrix $M_f$ provides a matrix representation of the linear operator ``multiplication by $f\brk{\xvec}$ modulo $\Ideal$'', expressed in the basis $\mons$ \cite{Sturmfels:2002,CoxLittleOshea:2005,Cox:2015ode,Telen:2020thesis}.
Companion matrices are a powerful tool as they form a commutative algebra: they are linear in the polynomial argument:
\begin{align}
    M_{f + g} = M_{f} + M_{g}
    \>,
\end{align}
they respect multiplication:
\begin{align}
    M_{f g} = M_{f} \cdot M_{g} = M_{g} \cdot M_{f}
    \>,
    \label{eq:cmat_product_rule}
\end{align}
and in particular they are pairwise commuting.
This property mirrors the commutativity of ordinary polynomial multiplication\footnote{
    The product rule~\labelcref{eq:cmat_product_rule} follows from application of the definition~\labelcref{eq:cmat_definition} to a product of polynomials $m_i \> f g$ twice: first to the inner product $m_i \> f = \sum_j \brk{M_{f}}_{ij} \> m_j \mod \Ideal$, and then to the outer product $m_j \> g = \sum_k \brk{M_{g}}_{jk} \> m_k \mod \Ideal$, which naturally gives rise to the matrix multiplication.
}.
Furthermore, assuming that the constant monomial $1$ appears in the basis $\mons$ in the rightmost position, the remainder of $f\brk{\xvec}$ can be straightforwardly extracted by contracting the $M_f$ matrix with the corresponding basis \brk{co}vector:
\begin{align}
    f\brk{\xvec} =
    \begin{bNiceMatrix}
        \mZero & \Ldots[color=gr] & \mZero & 1
    \end{bNiceMatrix}
    \cdot
    M_{f}
    \cdot
    \mons\tr
    \mod \Ideal
    \>.
\end{align}
To recover the polynomial remainder for $\textit{any}$ function $f(\xvec)$, it is thus sufficient to compute the companion matrices for each variable in $\Ideal$, collectively denoted as
\begin{align}
    M_\xvec = \{M_{x_1},\cdots, M_{x_\numvars} \}
    \>.
\end{align}
The construction of $M_f$ then requires no further polynomial division.

As an example, we consider once again $\Ideal$ given in~\cref{eq:ideal_2var_explicit}.
We have
\begin{equation}
    \left.
    \begin{aligned}
        y \cdot x &= y+1\>,\\
        1 \cdot x &=y+1\>,
    \end{aligned}
    \qquad
    \begin{aligned}
        y \cdot y &= 1\\
        1 \cdot y &=y
    \end{aligned}
    \quad \right\} \hspace{-8pt} \mod \Ideal\,,
\end{equation}
which can be computed via Gr\"obner bases (or Macaulay systems) as discussed above. The two companion matrices are thus given by
\begin{equation}
    M_x = 
    \begin{bNiceMatrix}
        1 & 1
        \\
        1 & 1
    \end{bNiceMatrix}
    \,,
    \quad
    M_y = 
    \begin{bNiceMatrix}
        \mZero & 1
        \\
        1 & \mZero
    \end{bNiceMatrix}
    \,.
\end{equation}
The companion matrix $M_f$ for $f(x,y)=x\,y^2$ can thus be computed as
\begin{equation}
    M_f = M_x \cdot M_y^2 =
    \begin{bNiceMatrix}
        1 & 1
        \\
        1 & 1
    \end{bNiceMatrix}
    \,,
\end{equation}
and the remainder can be extracted as
\begin{equation}
    f(x,y) = \begin{bNiceMatrix}
        \mZero & 1
    \end{bNiceMatrix}\cdot
    \begin{bNiceMatrix}
        1 & 1
        \\
        1 & 1
    \end{bNiceMatrix}
    \cdot
    \begin{bNiceMatrix}
        y
        \\
        1
    \end{bNiceMatrix}
    =
    y+1 \mod \Ideal\,.
\end{equation}
Likewise for $f(x,y) = x^{13} y^{27}$ we can write
\begin{equation}
    M_f = M_x^{13}\cdot M_y^{27} = 2^{12}
    \begin{bNiceMatrix}
        1 & 1
        \\
        1 & 1
    \end{bNiceMatrix}
    \,,
\end{equation}
and thus
\begin{equation}
    f(x,y) = 2^{12}\,(y+1)\mod\Ideal\,.
\end{equation}

\subsubsection{Rational Function Reduction}
\label{sec:cmat_rat_fun}
Companion matrices also extend to the reduction of rational functions.
In this case, the multiplicative inverse is represented by matrix inverse\footnote{
    This naturally follows from the multiplication property~\cref{eq:cmat_product_rule}: if two polynomials $f$ and $g$ are such that $f g = 1 \mod \Ideal$, then their companion matrices will satisfy $M_f \cdot M_g = \mId \mod \Ideal$, from which follows the property $M_g = M_f^{-1}$ whenever the matrix inverse exists.
},
\begin{align}
    M_{f_{\mathrm{inv}}} = M^{-1}_f\,.
    \label{eq:cmat_inverse_rule}
\end{align}
Thus, the algebra of companion matrices not only encodes the polynomial algebra, but can also naturally accommodate for rational functions.
For example,
\begin{equation}
    f(x,y)
    =
    \frac{x}{y^{100}-3x+2} + a
    \Longrightarrow M_{f}
    =
    M_x \left(M_y^{100} -3 M_x + 2 \, \mId \right)^{-1} + a \, \mId
    =
    \begin{bNiceMatrix}
        \frac{1}{3} (3 a-1) & -\frac{1}{3}
        \\
        -\frac{1}{3} & \frac{1}{3} (3 a-1)
    \end{bNiceMatrix}
    \,,
\end{equation}
and thus we can immediately deduce
\begin{equation}
    f(x,y) = \frac{1}{3} (3 a-1)-\frac{y}{3} \mod \Ideal\,.
\end{equation}

\subsubsection{Roots and the Eigenvalue Theorem}
\label{sec:eigenvalue_theorem}
There is an important connection between the properties of companion matrices and the algebraic structure of the ideal: the eigenvalues of the companion matrices $M_\xvec$ jointly encode the complete set of the roots of the ideal $\Ideal$.
Indeed, due to the commutativity property~\cref{eq:cmat_product_rule}, the companion matrices $M_{x_i}$ for each coordinate $x_i$ can be simultaneously diagonalized\footnote{
    This can always be done if the ideal $\Ideal$ is radical.
}.
Thus they share a common set of eigenvectors $\vvec\supbrk{\alpha}$ each individually associated with the roots $\alpha \in V\brk{\Ideal}$.
The corresponding eigenvalues $M_{x_i} \vvec\supbrk{\alpha} = \lambda_i \vvec\supbrk{\alpha}$ are precisely the $x_i$-coordinate values of the root $\alpha$, so that the vanishing locus is completely determined by the eigenvalues of the companion matrices:
\begin{align}
    V\brk{\Ideal} = \bigbrc{
        \brk{\lambda_1, \ldots, \lambda_\numvars} \in \Complex^\numvars
        \>\big|\>
        \text{there is $\vvec$ such that}
        \hspace{.7em}
        M_{x_i} \vvec = \lambda_i \vvec
        \hspace{.7em}
        \text{for every $i = 1, \ldots, \numvars$}
    }\,.
\end{align}
This statement is sometimes referred to as the \emph{Stickelberger} or the \emph{Eigenvalue Theorem}\footnote{
    For a proof of the theorem see, for example~\cite{Sturmfels:2002}, and for a historical review on its origins see~\cite{Cox:2020}.
}.
From practical point of view, this theorem allows one to study both numerically\footnote{
    The \spqr{} setup is flexible enough to construct companion matrices w.r.t. any non-degenerate basis. This is done by augmenting the Macaulay system with its defining equations and reordering the unknowns so that reduction proceeds in the new basis, analogous to the reduction of Feynman integrals via IBP methods.
    Alternative reduction bases may improve numerical stability in root determination problems, see, for example, the review in~\cite{Telen:2020thesis}.
} and symbolically the values of functions localized on solutions of systems of polynomial equations. Next we show one application of this kind.

\subsection{Elimination theory}
\label{sec:elimination_theory}
\begin{figure}[H]
    \centering
    \includegraphicsbox[width=0.3\textwidth]{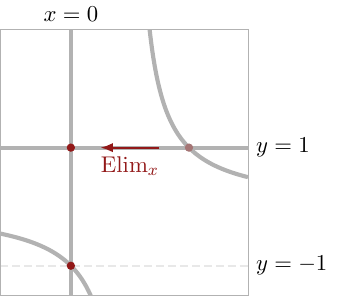}
    \hspace{1cm}
    \includegraphicsbox[width=0.3\textwidth]{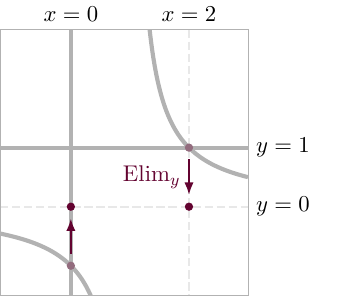}
    \caption{
        Projections of the root system from~\cref{eq:elim_2var_explicit} obtained by eliminating the variables $y$ \brk{left} and $x$ \brk{right}.
    }
    \label{fig:elimination_example}
\end{figure}

Elimination theory studies the problem of eliminating variables from polynomial systems, with the ultimate goal of solving polynomial systems of equations.
For example, one may be interested in reducing a system of multiple variables $\xvec$ onto a single equation in one variable $x$.
This reduction provides a pathway to analyse and solve the original system by focusing on simpler, univariate equations.
For example, taking $\Ideal = \langle xy -x, xy -y -1 \rangle \in \Rationals [x,y]$ as in \cref{eq:ideal_2var_explicit} we have
\begin{equation}\label{eq:elim_2var_explicit}
    \Elim_{x}(\Ideal) =  \langle (y-1)(y+1)\rangle=\Ideal \cap \Rationals[y],\qquad \Elim_{y}(\Ideal) = \langle x(x-2)\rangle = \Ideal \cap \Rationals[x]\,,
\end{equation}
which can be verified against the roots given in \cref{eq:ideal_explicit_roots}. Graphically, elimination can be interpreted as projecting the solutions of a polynomial system onto a lower dimensional subsystem, as illustrated in \cref{fig:elimination_example}.
\spqr{} provides three methods for eliminating variables from polynomial systems: via companion matrices and characteristic polynomials, via ansatz for companion matrices, and via elimination monomial orderings.
In the following sections, we discuss the necessary theoretical background pertaining to these three methods.
\subsubsection{Companion Matrix and Characteristic Polynomial Approach}\label{sec:charp_elim}
As overviewed in \cref{sec:companion_matrices}, the eigenvalues of a given companion matrix $M_x$ encode the $x$ coordinates of the vanishing set $V(\Ideal)$.
To find $\Elim_x(\Ideal)$, one could thus proceed by diagonalising $M_x$.
Whilst this approach is technically possible, it presents a similar complexity to directly solving for $\Ideal = 0$.
Instead, the root information can be implicitly extracted by building the characteristic polynomial $p_x(\lambda) := \det (M_x - \lambda \mathbb{1})$.

By definition $p$ vanishes precisely on the eigenvalues of its respective matrix.
It follows that $p_x$ must in turn vanish always and only on the $x$ coordinates of $V(\Ideal)$.
Thus\footnote{Technically speaking, the characteristic polynomial may produce factors with higher multiplicities than the true elimination ideal. Nevertheless the roots of both objects will always coincide.}
\begin{equation}
    \Elim_{\widehat{x}}(\Ideal) = p_x(x)=\det(M_x - x \mathbb{1})\,,
\end{equation}
where by $\widehat{x}$ we mean every variable except $x$. In cases where the characteristic polynomial factorises then one may want to discard any redundant multiplicities. This new object is known as the \emph{minimal polynomial} \cite[Corollary 4.6]{CoxLittleOshea:2005}.

An important advantage of eliminating variables with this approach is that it is independent of the monomial ordering used.
Thus, computing $M_{\xvec}$ and thus $p$ can be performed with the best ordering for a given ideal, which in many cases can significantly speed up computations.
To illustrate this method in action, we can derive \cref{eq:elim_2var_explicit}. We have
\begin{equation}\label{eq:elim_from_cmat_explicit}
\begin{aligned}
    p_y(y)=\Elim_x(\Ideal) &= \det(M_y-y\mathbb{1}) = \det
    \begin{bNiceMatrix}
        -y & 1
        \\
        1 & -y
    \end{bNiceMatrix}
    = (y-1)(y+1)\,,\\
    p_x(x)=\Elim_y(\Ideal)&=\det(M_x - x\mathbb{1}) = \det
    \begin{bNiceMatrix}
        1-x & 1
        \\
        1 & 1-x
    \end{bNiceMatrix}
    = x(x-2)\,,
\end{aligned}
\end{equation}
as before. The implementation of this procedure in \spqr{} is discussed in more detail in \cref{sec:theoretical_implementation}.
\subsubsection{Companion Matrix and Ansatz Approach}\label{sec:cmat_elim_ansatz}
Whilst many features of polynomial division depend on the given choice of monomial ordering, some important properties are true for any ordering. One such example is that of \emph{ideal membership}: if $f(\xvec) = 0 \mod \Ideal$ in one ordering, then $f = 0 \mod \Ideal$ for \emph{any} (valid) choice.

In terms of companion matrices, this implies that
\begin{equation}\label{eq:cmat_vanishing_cond}
    \begin{bNiceMatrix}
        \mZero & \Ldots[color=gr] & \mZero & 1
    \end{bNiceMatrix}
    \cdot
    M_{f}=\mathbf{0}\,,
\end{equation}
no matter the chosen monomial ordering.
This property can be used to eliminate variables from $\Ideal$: Suppose that $f \in \Ideal$, and that furthermore $f$ only depends on a (known) subset of the variables, $\yvec \in \xvec$ (in other words in $f$ the variables $\xvec \setminus \yvec$ have been eliminated).
An ansatz for $f$ can then be written as
\begin{equation}
    f(\yvec) = \sum_{\nvec} c_\nvec\,\yvec^{\nvec}\,,
\end{equation}
where the coefficients $c_\nvec$ are unknown and only finitely many of them are non-zero.
These however can straightforwardly be solved for by using \cref{eq:cmat_vanishing_cond},
\begin{equation}
    \begin{bNiceMatrix}
        \mZero & \Ldots[color=gr] & \mZero & 1
    \end{bNiceMatrix}
    \cdot
    M_{f(\yvec)} = 
    \begin{bNiceMatrix}
        \mZero & \Ldots[color=gr] & \mZero & 1
    \end{bNiceMatrix}
    \cdot
    \sum_{\nvec} c_\nvec\, 
    M_{\yvec}^{\nvec} = \mathbf{0}\,.
\end{equation}
The full form of $f$ can then be inferred this way. Thus variables can be eliminated from $\Ideal$ despite not having to use an elimination ordering in the construction of $M_{\xvec}$. This approach is very similar to \soft{FGLM} and related algorithms for converting Gr\"obner bases between each other \cite{FAUGERE2017538,FAUGERE1993329,Collart1997ConvertingBW}.

The key difference between this approach and that of \cref{sec:charp_elim} is that this method can be used to eliminate \emph{fewer} than all variables except one. Furthermore, this method will compute exactly the elimination ideal, which can manifest as lower factor multiplicities when compared to the characteristic polynomial approach. This in turn can result in a significant reduction in the required number of sample points in the reconstruction of the coefficients.

For this method to work properly an appropriate ansatz must first be found and provided. In practice for problems where \spqr{} is useful, obtaining this information is computationally manageable and does not pose serious computational bottlenecks. This is explained in more detail later in \cref{sec:advantages_and_disadvantages}.
\subsubsection{Elimination Order Approach}
\spqr{} also supports the more ``traditional" approach to eliminating variables, namely through the use of \textit{elimination monomial orders}.

Elimination orders are special types of monomial orderings designed to systematically remove certain variables.
Concretely, suppose the system in consideration has $\xvec$ variables and a subset $\yvec \in \xvec$ needs to be eliminated.
An elimination order is built such that \textit{every} monomial involving \textit{any} of the $\yvec$ variables is ranked higher than \textit{any} monomial involving only the other variables.
Lexicographic order is an example of an elimination order.
Indeed, \cref{eq:lexicographic_order} eliminates $x$ from $\{x,\,y\}$ as any power of $x$ is considered higher weight than any power of $y$: $x^i y^a > x^j y^b\,\, \text{if} \,\,i>j\,\, \forall\,\, a,b\,\, $.

By computing a Gr\"obner basis $\Groebner$ with respect to an elimination ordering, the eliminated system can immediately be read off as the new subset of generators no longer containing the eliminated variables\footnote{If such generators exist.}.
As an example, from $\Groebner$ in \cref{eq:multivar_example_gb}, we have that $g_1 = y^2-1$ does not contain $x$. Indeed, $y^2-1 = \Elim_x(\Ideal)$, as already shown in \cref{eq:elim_2var_explicit,eq:elim_from_cmat_explicit}.

The Gau\ss{}ian elimination strategy discussed in \cref{sec:multi_row_reduction} can also be used to compute $\Elim_x(\Ideal)$ without having to explicitly generate $\Groebner$: from the last row of \cref{eq:macaulay_2var_row_reduced} one can also immediately infer that $y^2 -1 = 0 \mod \Ideal$.%

Despite its conceptual simplicity, lexicographic order is often inefficient for variable elimination.
To address this, \spqr{} also supports block elimination orders, in which variables are grouped according to their roles, and monomials within each block are ordered using degree reverse lexicographic (or related) weighting. For example, the weight matrix corresponding to an elimination order of five variables, divided into groups of 2 and 3 elements, has the form
\begin{equation}
    W_{\mathrm{elim}} = \begin{bNiceMatrix}[margin, left-margin=6pt]
        \CodeBefore
            \rectanglecolor{red!20}{1-1}{2-2}
            \rectanglecolor{purple!20}{3-3}{6-6}
        \Body
            \color{red}{1} & \color{red}{1} & \mzero & \mzero & \mzero
            \\
            \mzeroRed & \color{red}{-1} & \mzero & \mzero & \mzero
            \\
            \mzero & \mzero & \color{purple}{1} & \color{purple}{1} & \color{purple}{1}
            \\
            \mzero & \mzero & \mzeroPurple & \mzeroPurple & \color{purple}{-1}
            \\
            \mzero & \mzero & \mzeroPurple & \color{purple}{-1} & \mzeroPurple
    \end{bNiceMatrix}
    \>.
\end{equation}
Reduction w.r.t. this order eliminates the
{\tikz[baseline] \node[rounded corners, fill=red!20, text=red, anchor=text] {red};}
variables in favor of the
{\tikz[baseline] \node[rounded corners, fill=purple!20, text=purple, anchor=text] {purple};}
ones.
\subsection{Implementation}\label{sec:theoretical_implementation}

In the previous sections, we have discussed the necessary theoretical background pertaining to polynomial division, as well as various related algorithms and operations that can be performed with this technology. In this section, we focus on how these ideas can be efficiently implemented in a computer, and how this is done specifically inside \spqr{}.

\subsubsection{Review of Finite Field Sampling and Reconstruction}
One of the major challenges in computer algebra is \emph{expression swell}: as symbolic manipulations proceed, intermediate expressions often become dramatically larger than the final simplified result.

A particularly effective way to circumvent this problem is through \emph{black-box rational reconstruction} combined with \emph{finite-field sampling}~\cite{Kauers:2008zz,Kant:2013vta,vonManteuffel:2014ixa,Peraro:2016wsq,Peraro:2019svx,Smirnov:2019qkx,Klappert:2019emp}.

First introduced to high energy physics via integration-by-parts reductions, this method has since become a central component of many modern computational pipelines in perturbative Quantum Field Theory.
The strategy is to reformulate the problem so that the desired quantities are represented as multivariate rational functions.
One then sets up an Ansatz with unknown coefficients and fixes the them by probing the system numerically as a black box at multiple sampling points. The exact rational result is subsequently reconstructed from several modular evaluations using the Chinese remainder theorem in combination with the Wang algorithm~\cite{Wang:1981, Wang:1982}.

This approach presents three important advantages. Firstly, instead of working with exact rational numbers $\Rationals$ during sampling, one evaluates the problem in the more efficient arithmetic setting of $\Integers_p$, with $p$ a large prime. Such finite-field arithmetic naturally caps the maximum expression size, avoiding expression swell even on numerical slices. Secondly, any complicated cancellations in the algorithm's output happen numerically, and before any interpolation/reconstruction. This effectively avoids the need for heavy intermediate manipulations required in a fully algebraic approach. Finally this strategy lends well to paralellisation, as numerical evaluations of sample points are fully independent calculations.

\subsubsection{Finite Field Sampling in \spqr{}}

\spqr{} makes extensive use of finite field sampling and reconstruction. We employ as a back end the high-performance \soft{C++} library
\soft{FiniteFlow}~\cite{Peraro:2019svx}, which provides a robust and flexible implementation of these strategies. \soft{FiniteFlow}'s design allows for a wide range of end user applications: in particular it naturally offers support for list and matrix manipulations. As outlined in the previous sections, these basic algebraic operations can be made to form the building blocks of all necessary operations for polynomial division. \spqr{} provides a high level user interface which automatically implements these processes inside \soft{FiniteFlow}, without requiring any knowledge from the user on the operation of the back end algorithms.

\begin{figure}[H]
    \centering
    \includegraphicsbox[width=0.4\textwidth]{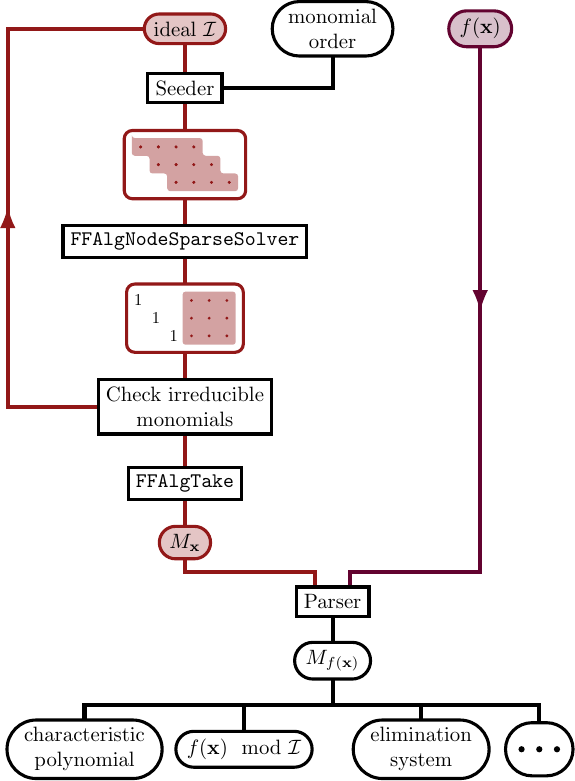}
    \caption{A flowchart showing the various stages in a computation in \spqr{}, where the various \soft{FiniteFlow} commands used as a back end have been labelled.}
    \label{fig:cmat_flowchart}
\end{figure}

Crucially, the program is specifically designed so that only the final quantities of direct interest are reconstructed. All intermediate steps, including Gau\ss{}ian elimination, companion matrix generation, companion matrix multiplication and characteristic polynomial construction are performed entirely numerically within \soft{FiniteFlow}. Importantly, this implies that \spqr{} never requires or builds an explicit form of the Gr\"obner basis, as this information is indirectly contained in the Macaulay system after row reduction has been performed\footnote{If the desired output of a computation \textit{is} a Gr\"obner basis, it is possible to explicitly reconstruct it with \spqr{}.}. Once the desired quantity has been built numerically, \spqr{} then reconstructs its full functional dependence. A schematic overview of \spqr{}'s internal workings is given in \cref{fig:cmat_flowchart}.
\subsubsection{Which Ring Does \spqr{} Use?}\label{sec:which_ring}
Due to its finite field back end, it is important to clarify the polynomial ring in which \spqr{} performs polynomial reduction. To do this, one must distinguish between \textit{variables} and \textit{parameters} in a polynomial ideal $\Ideal$. The variables are the objects with respect to which polynomial reduction takes place. In contrast, the parameters appear only in the coefficients of the polynomials, and each coefficient can be a rational function in the parameters. For instance, in the univariate example of \cref{eq:univ_example}, the variable is $x$ and the parameter is $a$. In the multivariate ideal of \cref{eq:ideal_2var_explicit}, the variables are ${x, y}$, and there are no parameters. 

By nature of finite field reconstruction algorithms, \spqr{} always operates in the ring
\begin{equation}
    R = \mathbb{Q}(\text{parameters})[\text{variables}]\,.
\end{equation}
In other words, the output of any \spqr{} computation will be (a set of) polynomials in the variables, with coefficients rational functions in the parameters. Indeed \spqr{} performs rational reconstruction only in the parameters — all the variables, upon building the Macaulay system, are contained in the external vector which never explicitly enters in any computations. Two examples of this in action are \cref{eq:1var_macaulay_matrix,eq:2var_macaulay_matrix}.
\subsubsection{Advantages and Disadvantages of \spqr{}'s Approach}\label{sec:advantages_and_disadvantages}
By design \spqr{} performs polynomial divisions in a different manner to most computer algebra systems. Whilst in many cases this can provide large computational benefit, there are also many examples where a more traditional computer algebra approach is more advantageous. Which algorithm performs best in each situation depends almost entirely by the structure of the ideal being considered.

Crudely speaking, the complexity of a polynomial ideal can be divided into two categories: ``variable complexity" and ``parameter complexity". Whilst seemingly similar, this distinction can enormously impact the efficiency of the chosen algorithm.

If a given ideal has many variables raised to high powers, then this will result in a difficult Gr\"obner basis computation with many intermediate steps. In \spqr{}, this translates to having to generate a large Macaulay system to high degree. By its nature, the Gau\ss{}ian elimination performed by \spqr{} is a cruder operation when compared to fine tuned Gr\"obner basis algorithms \cite{FAUGERE199961}. Thus, for ideals with complicated variable dependence (and no or few parameters to reconstruct) it is more likely that traditional polynomial division strategies will outperform \spqr{}.

Conversely, there exist
many ideals where the variable complexity is contained, but there are many additional parameters in the polynomial system. In these cases, even if the Gr\"obner basis requires relatively few operations to calculate, symbolic computer algebra approaches may suffer greatly due to intermediate expression swell. Due to its finite field back end, \spqr{} does not suffer from this problem and thus can efficiently handle ideals with very complicated parameter dependence.

In summary, the worst case scenario for \spqr{} is an ideal in many variables with high powers and no parameters. The best case scenario instead is an ideal that is relatively simple on a numerical slice, but has many parameters creating intermediate expression swell, hindering the efficiency of algebraic algorithms when trying to work in the full parameter space.

Fortunately, many problems in high energy physics and beyond fall into this latter category, where (physical) parameters are almost universal. For such systems, we find \spqr{}'s approach to enormously beneficial compared to algebraic approaches. A more quantitative analysis of specific relevant examples is provided in \cref{sec:examples}.
\subsubsection{Further Implementation Details}
\label{sec:further_details}
\begin{figure}[H]
    \centering
    \includegraphicsbox[width=0.4\textwidth]{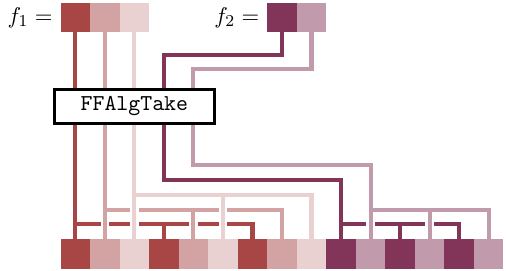}
    \caption{
        Structure of non-zero elements of the Macaulay system from~\cref{eq:system_generation_markandsweep}.
        The \code{FFAlgTake} function of \soft{FiniteFlow} rearranges the coefficiants of the input polynomials $f_1$ and $f_2$ \brk{shown in the top row} into the non-zero entries of the sparse Macaulay matrix, stored in row-major order.
    }
    \label{fig:macaulay_nonzero}
\end{figure}

\paragraph{System generation}
The first step of our method is the construction of the linear Macaulay system that encodes the algebraic problem.
\spqr{} uses a straightforward seeding strategy: each polynomial in $\Ideal$ is multiplied by all monomials up to a given total degree $d$, where $d({x_1}^{n_1}\,{x_2}^{n_2}\ldots) = n_1 + n_2 + \cdots$.
The system is generated directly within \soft{FiniteFlow}, which we find significantly improves performance at this stage of the algorithm.
This is achieved by exploiting the sparsity of the Macaulay matrix to only construct the non-zero entries from the coefficient of the generator polynomials, as illustrated in~\cref{fig:macaulay_nonzero} for the example discussed in~\cref{sec:multi_row_reduction}.

As already mentioned in \cref{sec:multi_row_reduction}, it is unclear a priori to what weight the Macaulay system must be generated in order to obtain a correct polynomial reduction. To ensure the correct result is always reached, \spqr{} provides tools to evaluate a Gr\"obner basis for $\Ideal$ once on a numerical slice\footnote{In practice, for problems where \spqr{} is useful, the computation of a numerical Gr\"obner basis is not a bottleneck, even with \soft{Mathematica}'s built in tools. Indeed as outlined in \cref{sec:advantages_and_disadvantages}, \spqr{} is most effective when the variable complexity of the ideal is contained. This is equivalent to Gr\"obner bases on numerical slices being accessible.}. From this information, the set of irreducible monomials can be extracted and compared against those found by the Gau\ss{}ian elimination strategy. If a match is found, then the seeding weight was sufficient. If not, larger systems can be iteratively generated until a sufficient weight is reached.

Finally, in most cases, the seeding strategy described above produces an overdetermined linear system, containing many more equations than are strictly necessary to determine all unknowns.
Consequently, it is highly desirable to find equivalent but smaller reformulations that preserve the algebraic content of the problem while substantially reducing computational cost.
Within \spqr{}, this simplification is achieved through \soft{FiniteFlow}'s Mark-and-Sweep algorithm, which automatically discards the redundant relations after solving the system numerically once.
This strategy is conceptually equivalent to the \textit{tracing} algorithms used by modern Gr\"obner basis implementations \cite{10.1007/3-540-51084-2_12}.
For the example discussed in~\cref{sec:multi_row_reduction}, the initial and truncated Macaulay systems take the following shape:

\newcommand{\RdotI}{}
\newcommand{\RdotII}{}
\newcommand{\RdotIII}{}
\newcommand{\PdotI}{}
\newcommand{\PdotII}{}

\begin{equation}
    \begin{bNiceMatrix}[margin, columns-width = 3mm]
        \CodeBefore
            \cellcolor{red1}{2-2, 3-4, 4-5}
            \cellcolor{red2}{2-5, 3-7, 4-8}
            \cellcolor{red3}{2-6, 3-8, 4-9}
            \cellcolor{purple1}{5-2, 6-4, 7-5}
            \cellcolor{purple2}{5-3, 6-5, 7-6}
        \Body
        1 & \mzero & \mzero & \color{gr5}{\bigcdot} & \mzero & \mzero & \mzero & \mzero & \mzero
        \\
        \mzero & \RdotI & \mzero & \mzero & \RdotII & \RdotIII & \mzero & \mzero & \mzero
        \\
        \mzero & \mzero & \mzero & \RdotI & \mzero & \mzero & \RdotII & \RdotIII & \mzero
        \\
        \mzero & \mzero & \mzero & \mzero & \RdotI & \mzero & \mzero & \RdotII & \RdotIII
        \\
        \mzero & \PdotI & \PdotII & \mzero & \mzero & \mzero & \mzero & \mzero & \mzero
        \\
        \mzero & \mzero & \mzero & \PdotI & \PdotII & \mzero & \mzero & \mzero & \mzero
        \\
        \mzero & \mzero & \mzero & \mzero & \PdotI & \PdotII & \mzero & \mzero & \mzero
        \CodeAfter
            \tikz \draw[
                line width=.4pt, gr,
            ]
                (1|-2) -- (10|-2)
                (1-|2) -- (9-|2)
            ;
    \end{bNiceMatrix}
    \xrightarrow{\texttt{FFSparseSolverMarkAndSweepEqs}}
    \begin{bNiceMatrix}[margin, columns-width = 3mm]
        \CodeBefore
            \cellcolor{red1}{2-5}
            \cellcolor{red2}{2-8}
            \cellcolor{red3}{2-9}
            \cellcolor{purple1}{3-4, 4-5}
            \cellcolor{purple2}{3-5, 4-6}
        \Body
        1 & \mzero & \mzero & \textcolor{gr5}{\bigcdot} & \mzero & \mzero & \mzero & \mzero & \mzero
        \\
        \mzero & \mzero & \mzero & \mzero & \RdotI & \mzero & \mzero & \RdotII & \RdotIII
        \\
        \mzero & \mzero & \mzero & \PdotI & \PdotII & \mzero & \mzero & \mzero & \mzero
        \\
        \mzero & \mzero & \mzero & \mzero & \PdotI & \PdotII & \mzero & \mzero & \mzero
        \CodeAfter
            \tikz \draw[
                line width=.4pt, gr,
            ]
                (1|-2) -- (10|-2)
                (1-|2) -- (6-|2)
            ;
    \end{bNiceMatrix}
    \label{eq:system_generation_markandsweep}
    \>,
\end{equation}
where the sparse matrix on the left is built from the non-zero entries illustrated in the bottom row of~\cref{fig:macaulay_nonzero}.

\begin{figure}[h]
    \centering
    \begin{subfigure}[t]{0.4\textwidth}
        \centering
        \includegraphicsbox[height=5cm]{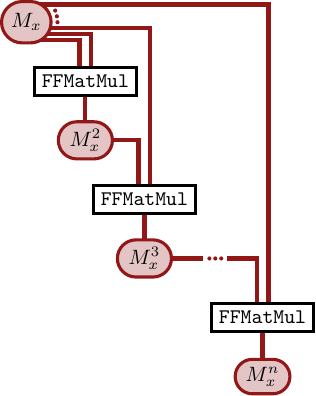}
        \subcaption{\texttt{MatPower[n]}}
        \label{fig:MatPower}
    \end{subfigure}
    \begin{subfigure}[t]{0.4\textwidth}
        \centering
        \includegraphicsbox[height=5cm]{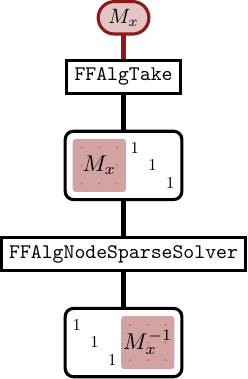}
        \subcaption{\texttt{MatInverse}}
        \label{fig:MatInverse}
    \end{subfigure}
    \caption{
        Flowcharts illustrating two matrix operations in \spqr{}: a simple recursive algorithm that computes the $n^{\text{th}}$ matrix power~\brk{\subref{fig:MatPower}}, and a subroutine for matrix inversion using a linear solver~\brk{\subref{fig:MatInverse}}.
    }
    \label{fig:MatPower_MatInverse}
\end{figure}

\begin{figure}[h]
    \centering
    \includegraphicsbox[height = 5cm]{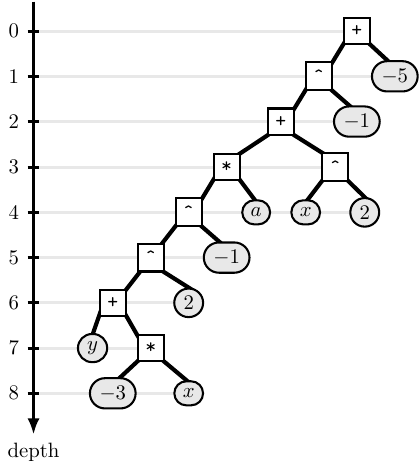}
    \hspace{1cm}
    \includegraphicsbox[height = 5cm]{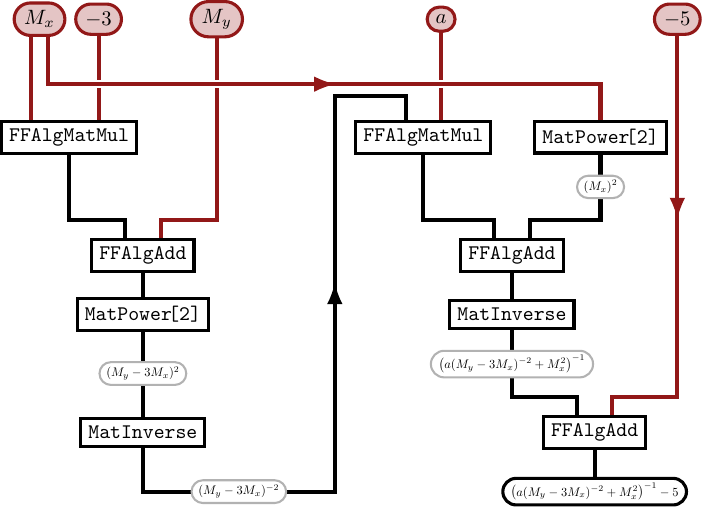}
    \caption{
        Expression tree of the rational function from~\cref{eq:ex_code_target} as revealed by \soft{Mathematica} command~\code{TreeForm} on the left and its automatically generated internal representation in \spqr{} on the right.
    }
    \label{fig:expression_tree}
\end{figure}

\paragraph{Expression parsing}
As discussed in \cref{sec:companion_matrices}, the reduction of complicated rational functions $f(\xvec)$ relies on the usage of companion matrices.
In \spqr{}, this process is handled by a recursive parser that automatically converts any rational function of arbitrary nested depth into its corresponding companion matrix representation within \soft{FiniteFlow}.

Each function $f\brk{\xvec}$ is represented as an \emph{expression tree}, whose leaves correspond to variables $\xvec$ and constants, while its internal nodes represent basic algebraic operations of addition, multiplication, and exponentiation $\brc{\text{\code{+}}, \text{\code{*}}, \text{\code{\textasciicircum}}}$. 
The parser traverses this tree recursively, replacing each algebraic operation with its matrix analogue and each leaf with the companion matrix representation of the corresponding variable or constant.

Whilst some of these required operations to parse rational functions are already present in \soft{FiniteFlow}, others are instead built from more basic operations. Two such cases are the recursive implementation of matrix powers, as well as the computation of matrix inverses, as illustrated in~\cref{fig:MatPower_MatInverse}.

An example of the automatically generated internal \spqr{} representation of the rational function from~\cref{eq:ex_code_target} is shown in~\cref{fig:expression_tree}.

\begin{figure}[H]
    \centering
    \includegraphicsbox[width=0.30\textwidth]{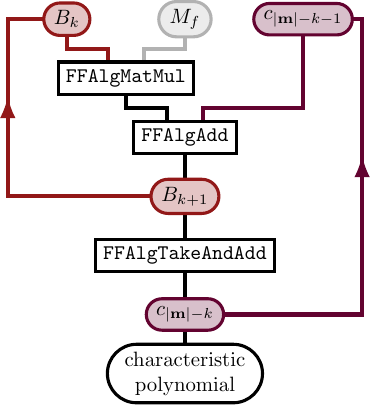}
    \caption{A flowchart showing \spqr{}'s implementation of the Faddeev–LeVerrier algorithm, where the various \soft{FiniteFlow} commands have been labelled.}
    \label{fig:char_p_algorithm}
\end{figure}

\paragraph{Characteristic polynomial algorithm}
To build the characteristic polynomials of companion matrices as discussed in \cref{sec:elimination_theory}, \spqr{} implements the Faddeev–LeVerrier algorithm. For a given $\numirreds \times \numirreds$ matrix this approach computes each coefficient of the characteristic polynomial 
\begin{equation}
    \det(M_f-\lambda \mathbb{1}) = c_0 + c_1 \lambda + \cdots + c_{\numirreds-1}\lambda^{\numirreds-1}+\lambda^{\numirreds}\,,
\end{equation}
without ever needing to explicitly introduce the eigenvalue parameter $\lambda$. The coefficients $c_{\numirreds-i}$ are computed recursively by introducing an auxiliary set of matrices $B_i$. The induction begins as
\begin{equation}
    B_0 = 0\,, \qquad \qquad  c_{\numirreds} = 1\,,
\end{equation}
and all following coefficients are computed with
\begin{equation}
    B_k = M_f\,B_{k-1} + c_{\numirreds-k+1} \mathbbm{1}\,, \qquad \qquad   c_{\numirreds-k} = -\frac 1 k \mathrm{tr}(M_f\,B_k)\,.
\end{equation}
The operations required to run this recursion are matrix traces and multiplications, which are both supported inside \soft{FiniteFlow}. In practice, we find this this algorithm to be very efficient, adding negligible computation time when compared to the (very often) more expensive previous row reduction steps. A schematic overview of the \spqr{}'s implementation showing the relevant \soft{FiniteFlow} functions is given in \cref{fig:char_p_algorithm}.

%% file: sections/program_description.tex
\newpage
\section{Program Installation and Usage}\label{sec:prog_install}
\subsection{Installation}
\spqr{} requires \soft{Mathematica} 13.1+ as well as the package \soft{FiniteFlow} \cite{Peraro:2019svx}, the repository for which can be found at: \href{https://github.com/peraro/finiteflow}{https://github.com/peraro/finiteflow}. 

With these prerequisites satisfied, \spqr{} can automatically be installed (or updated) with the command:
\begin{lstlisting}
ResourceFunction["GitHubInstall"]["giu989","SPQR"];
\end{lstlisting}
which can be run from any \soft{Mathematica} notebook or kernel session. This command will download all relevant files and install them along with \spqr{}'s built in documentation\footnote{We note that \soft{Mathematica} may need to be closed and reopened for the documentation files to be installed correctly.}. The package should from then on be loadable as usual with:
\begin{lstlisting}
<<SPQR`
\end{lstlisting}
If preferred, the source code as well as instructions for manual installation can be found on the \spqr{} GitHub page: \href{https://github.com/giu989/spqr}{https://github.com/giu989/spqr}.

To uninstall \spqr{}, run the command:
\begin{lstlisting}
PacletUninstall["SPQR"];
\end{lstlisting}

\subsection{Quickstart guide}
\begin{figure}[H]
    \centering
        \begin{subfigure}[t]{0.32\textwidth}
        \centering
        \includegraphicsbox[scale=0.7]{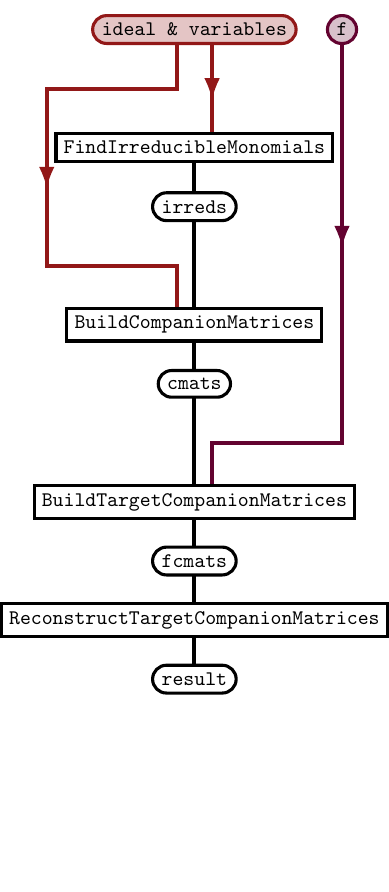}
        \caption{}
        \label{fig:quickstart_polydiv}
    \end{subfigure}
    \begin{subfigure}[t]{0.32\textwidth}
        \centering
        \includegraphicsbox[scale=0.7]{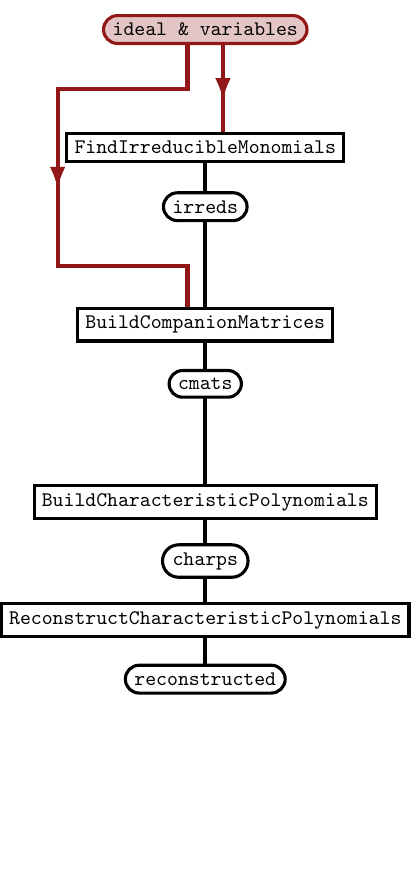}
        \caption{}
        \label{fig:quickstart_charpoly}
    \end{subfigure}
    \begin{subfigure}[t]{0.32\textwidth}
        \centering
        \includegraphicsbox[scale=0.7]{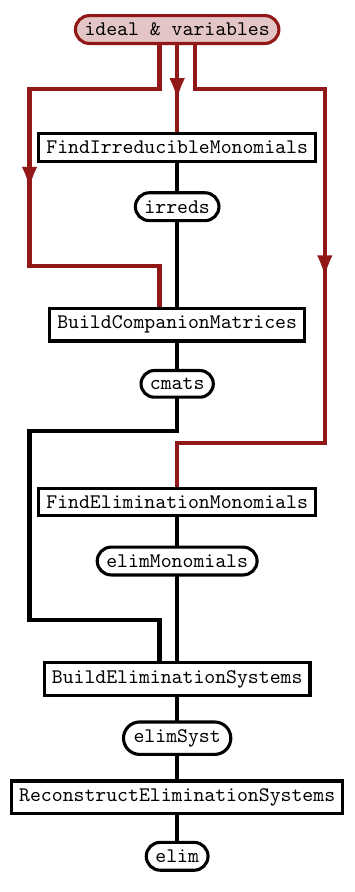}
        \caption{}
        \label{fig:quickstart_multivar}
    \end{subfigure}
    \caption{
        Flowcharts of the three main computational pipelines in \spqr{}: division of multivariate rational functions \brk{\subref{fig:quickstart_polydiv}}, elimination of all variables but one using characteristic polynomials \brk{\subref{fig:quickstart_charpoly}}, and elimination of fewer than all variables but one based on companion matrix ansatzes \brk{\subref{fig:quickstart_multivar}}.
    }
    \label{fig:quickstart}
\end{figure}

In this section we illustrate the most important workflows inside \spqr{} applied to some simple examples. Specifically we discuss show how \spqr{} can be used to perform polynomial division of rational functions as well as two methods for eliminating variables. Each one of these procedures can be thought of as a computational ``pipeline" where various \spqr{} routines are called in specific orders. These are summarised in the flowcharts in \cref{fig:quickstart}

Please note that a more in depth, interactive tutorial along with detailed descriptions and options for each function is automatically installed into \spqr{}. These files can be easily accessed upon loading the package via the dedicated ``Open documentation" button, or for example with the command \code{?SPQRDet}.

\subsubsection{Polynomial Division with Companion Matrices}
\label{sec:quickstart_polydiv}
\paragraph{First Example}
We begin by showing how \spqr{} can be used for polynomial division via companion matrices (see~\cref{fig:quickstart_polydiv} for a flowchart displaying the various used commands). We define an ideal in two variables $x,y$ 
as well as a parameter $a$,
\begin{equation}\label{eq:ex_code_ideal}
    \Ideal = \left\langle a\, x^2 y^2-2 x^2+3,-2 x\,y+y^2-3 y+1\right\rangle\,,
\end{equation}
as well as two polynomials and a rational function to reduce
\begin{equation}\label{eq:ex_code_target}
    \fvec = \left\{a\,x^2 + y^3 + y^2 + 3 + x\,y^2, a + x\,y^2,\frac{1}{\frac{a}{(y-3 x)^2}+x^2}-5\right\}.
\end{equation}
In \soft{Mathematica} they are written as
\begin{lstlisting}
variables = {x, y};
ideal = {a*y^2*x^2 - 2 x^2 + 3, y^2 - 3 y + 1 - 2*x*y};
f = {a*x^2 + y^3 + y^2 + 3 + x*y^2, a + x*y^2,-5 + (x^2 + a/(-3*x + y)^2)^(-1)};
\end{lstlisting}
As a first step, we require to identify a basis of irreducible monomials for this ideal. This can be done with the \spqr{} command:
\begin{lstlisting}
irreds = FindIrreducibleMonomials[ideal, variables]
(*{y^5, y^4, y^3, y^2, y, 1}*)
\end{lstlisting}
To keep convention with \soft{Mathematica}, by default lexicographic ordering is assumed. The next stage in the pipeline is to build the companion matrices $M_x$ and $M_y$ from this ideal. This is done by calling:
\begin{lstlisting}
cmats = BuildCompanionMatrices[ideal, variables, {1, 10}, irreds];
\end{lstlisting}
The third input, \code{\{1,10\}}, specifies the minimum and maximum weight to which the Macaulay system must be generated. \spqr{} will start at the lowest weight specified and iteratively build a larger matrix until the system is large enough, or the maximum is reached. For this example, weight three is sufficient. The next step is to build the companion matrices $M_\fvec$ for the target polynomials with:
\begin{lstlisting}
fcmats = BuildTargetCompanionMatrices[f, cmats];
\end{lstlisting}
Finally, the result of polynomial division is reconstructed with the command:
\begin{lstlisting}
result = ReconstructTargetCompanionMatrices[fcmats];
\end{lstlisting}
Explicitly, the answer for the first two entries reads
{
\setlength{\extrarowheight}{2mm}
\begin{equation}
    \code{result[[1;;2]]}=
    \begin{bNiceMatrix}
     \frac{a^2 y^4}{8}-\frac{3}{4} \left(a^2-2\right) y^3+\frac{1}{8} \left(11 a^2-4\right) y^2+\frac{1}{4} \left(2-3 a^2\right) y+\frac{1}{8} \left(a^2+12 a+24\right)
        \\
     a+\frac{y^3}{2}-\frac{3 y^2}{2}+\frac{y}{2}
    \end{bNiceMatrix}
    \,,
\end{equation}
}
which can be checked against \soft{Mathematica's} built in functions with:
\begin{lstlisting}
gb = GroebnerBasis[ideal, variables, CoefficientDomain->RationalFunctions];
gbAns = PolynomialReduce[f[[1;;2]], gb, variables] // Map[Last];
result[[1;;2]] - gbAns // Factor
(*{0,0}*)
\end{lstlisting}
\soft{Mathematica} does not have built in functionality for multivariate polynomial division of rational functions. Nevertheless, the division of the 3rd entry can be checked by using \cref{eq:univ_inverse_def,eq:univ_rat_reduction} with the following code:
\begin{lstlisting}
t1 = PolynomialReduce[result[[3]]*(f[[3]] // Together // Denominator),gb, variables][[2]];
t2 = PolynomialReduce[(f[[3]] // Together // Numerator), gb, variables][[2]] // Factor;
t1 == t2 // Factor
(*True*)
\end{lstlisting}
We note that \spqr{} did not need to generate an explicit Gr\"obner basis to obtain the correct reductions.
\paragraph{Monomial Orders}
Often one is interested in using monomial orders other than lexicographic. This can be specified with the option \code{"MonomialOrder"}, which needs to be passed to the commands \code{FindIrreducibleMonomials} and \code{BuildCompanionMatrices} as follows:

\begin{minipage}{\linewidth}
\begin{lstlisting}
(*pass the option in these commands*)
irreds = FindIrreducibleMonomials[ideal, variables, 
        "MonomialOrder"->DegreeReverseLexicographic];
cmats = BuildCompanionMatrices[ideal, variables, {1, 10}, irreds, 
        "MonomialOrder"->DegreeReverseLexicographic];
\end{lstlisting}
\end{minipage}
\begin{lstlisting}
(*the rest do not change*)
fcmats = BuildTargetCompanionMatrices[f, cmats];
result = ReconstructTargetCompanionMatrices[fcmats];
\end{lstlisting}
User defined weight matrices can also be accepted as monomial orders. For a description of all function options, see \cref{sec:description_of_all_functions} or \spqr{}'s built in documentation.

\subsubsection{Eliminating Variables with Characteristic Polynomials}\label{sec:var_elim_w_cmat}
We now turn to showing how \spqr{} can be used to eliminate variables from a system of equations using companion matrices \brk{see~\cref{fig:quickstart_charpoly} for a flowchart}. We take the ideal given in \cref{eq:ex_code_ideal}, and ask to eliminate the variable $y$. We begin by setting up the problem:
\begin{lstlisting}
variables = {x, y};
ideal = {a*y^2*x^2 - 2 x^2 + 3, y^2 - 3 y + 1 - 2*x*y};
\end{lstlisting}
As before, we find the irreducible monomials and build the companion matrices $M_x$ and $M_y$ for the ideal:
\begin{lstlisting}
irreds = FindIrreducibleMonomials[ideal, variables, 
        "MonomialOrder"->DegreeReverseLexicographic];
cmats = BuildCompanionMatrices[ideal, variables, {1, 10}, irreds, 
        "MonomialOrder"->DegreeReverseLexicographic];
\end{lstlisting}
This computation is monomial order independent, which is why (the in general more efficient) \newline \code{DegreeReverseLexicographic} ordering is chosen here. Next, the characteristic polynomial for $M_x$ is built with:
\begin{lstlisting}
charps = BuildCharacteristicPolynomials[cmats, {1}];
\end{lstlisting}
where the second argument tells \spqr{} to only compute the characteristic polynomial for the first entry in \code{variables}, $x$. Finally, the result is reconstructed using the command:
\begin{lstlisting}
reconstructed = ReconstructCharacteristicPolynomials[charps]
(*{{-(9/(8 a)), 0, (3/2 - (21 a)/8)/a, -(9/2), (-(1/2) + a/4 -a^2/8)/a, 3, 1}}*)
\end{lstlisting}
\code{ReconstructCharacteristicPolynomials} returns the list of coefficients $c_i$ of the characteristic polynomial (with normalisation $c_n=1$). To restore the explicit dependence on $x$ this can be achieved with a final algebraic post processing step:
\begin{lstlisting}
elim = reconstructed[[1]].(x^(Range[reconstructed[[1]]//Length]-1)) // Together // Numerator;
\end{lstlisting}
This agrees (up to an irrelevant overall sign) with \soft{Mathematica's} built in elimination tools:

\begin{minipage}{\linewidth}
\begin{lstlisting}
gbElim = GroebnerBasis[ideal, {x}, {y}, CoefficientDomain -> RationalFunctions] // First;
elim/gbElim // Factor
(*-1*)
\end{lstlisting}
\end{minipage}
\subsubsection{Eliminating Variables via Companion Matrix Ansatz}\label{sec:elim_cmat_ansatz}
Here we show how \spqr{} can be used to eliminate variables via the ansatz method discussed in \cref{sec:cmat_elim_ansatz} \brk{see~\cref{fig:quickstart_multivar} for a flowchart}. We begin by defining a zero-dimensional ideal in three variables $\{x,y,z\}$ as well as two parameters $\{a,b\}$,
\begin{equation}
    \Ideal = \langle -3\, a+x+y+z,b\, x^2\, y\, z-5,x\, y\, z-3\, z+3 \rangle\,.
\end{equation}
The goal will be to eliminate $\{z\}$ from this system. In \soft{Mathematica} this information is written as:
\begin{lstlisting}
vars = {z, y, x};
ideal = {x + y + z - 3 a, b*x^2*z*y - 5, x*y*z - 3 z + 3};
\end{lstlisting}
As before a set of irreducible monomials and companion matrices are computed:
\begin{lstlisting}
irreds = FindIrreducibleMonomials[ideal, vars, "MonomialOrder" -> DegreeReverseLexicographic]
(*{y, x, 1}*)
cmats = BuildCompanionMatrices[ideal, vars, {1, 10}, irreds,
        "MonomialOrder"->DegreeReverseLexicographic];
\end{lstlisting}
The computation does not depend on the monomial order of the companion matrices, which is why the generally more favourable \code{DegreeReverseLexicographic} was chosen above. Next, the monomials appearing in the eliminated ideal are found with:
\begin{lstlisting}
elimMonomials = FindEliminationMonomials[ideal, {z}, {y, x}]
(*{{x^3, x^2, x, 1}, {y, x^2, x, 1}}*)
\end{lstlisting}
Finally, the ansatz is built with:
\begin{lstlisting}
elimSyst = BuildEliminationSystems[cmats, elimMonomials];
\end{lstlisting}
and the result reconstructed by using:
\begin{lstlisting}
elim = ReconstructEliminationSystems[elimSyst];
\end{lstlisting}
which can be checked against \soft{Mathematica}'s built in elimination tools with
\begin{lstlisting}
elimgb = GroebnerBasis[ideal, {y, x}, {z}, CoefficientDomain -> RationalFunctions];
(elim // Together // Numerator)/elimgb // Factor
(*{-1,-1}*)
\end{lstlisting}

\subsubsection{Polynomial Division Without Companion Matrices}
\spqr{} also supports directly computing polynomial remainders from Macaulay systems, without having to pass through intermediate companion matrices. To illustrate how this works, we can consider the setup already discussed in \cref{eq:ex_code_ideal,eq:ex_code_target}. Explicitly we have once again:

\begin{minipage}{\linewidth}
\begin{lstlisting}
variables = {x, y};
ideal = {a*y^2*x^2 - 2 x^2 + 3, y^2 - 3 y + 1 - 2*x*y};
f = {a*x^2 + y^3 + y^2 + 3 + x*y^2, a + x*y^2};
\end{lstlisting}
\end{minipage}

as well as:
\begin{lstlisting}
irreds = FindIrreducibleMonomials[ideal, variables]
(*{y^5, y^4, y^3, y^2, y, 1}*)
\end{lstlisting}
We can then generate a Macaulay system directly with the command
\begin{lstlisting}
system = BuildPolynomialSystem[f, ideal, variables, {1, 10},
        "IrreducibleMonomials" -> irreds];
\end{lstlisting}
The result can then be reconstructed with:
\begin{lstlisting}
result = ReconstructPolynomialRemainder[system];
\end{lstlisting}
which can once again be double checked by running:
\begin{lstlisting}
gb = GroebnerBasis[ideal, variables, CoefficientDomain->RationalFunctions];
gbAns = PolynomialReduce[f, gb, variables] // Map[Last];
result - gbAns // Factor
(*{0,0}*)
\end{lstlisting}
For complicated cases, where polynomials with high powers need to be reduced, the size of the Macaulay system which needs to be generated can become very large with this approach. For this reason for the majority of cases we recommend using the companion matrix pipeline shown above instead.
\subsection{Description of all Functions}\label{sec:description_of_all_functions}
For completeness, in this section we provide descriptions of all functions in \spqr{} as well as tabulate their options. For more detailed usage, including examples for each option value, we recommend to read \spqr{}'s built in documentation.
\subsubsection{FindIrreducibleMonomials}\label{sec:findirmons}
\code{FindIrreducibleMonomials[ideal,vars]}
finds the irreducible monomials of an \code{ideal} in the variables \code{vars} using a numerical Groebner Basis.
\begin{center}
        \begin{tabular}{|l|l|l|}
        \hline
        \multicolumn{3}{|c|}{\textbf{Options for} \code{FindIrreducibleMonomials}} \\
        \hline
        \textbf{Option} & \textbf{Default Value} & \textbf{Description} \\
        \hline
        \texttt{"MonomialOrder"} & \texttt{Lexicographic} & monomial order to use \\
        \texttt{"Sort"} & \texttt{False} & attempts to find a more optimal ordering \\
        \hline
        \end{tabular}
\end{center}
The code for this function was adapted from \cite{Crisanti:2025gcs}.
\subsubsection{BuildCompanionMatrices}
\code{BuildCompanionMatrices[ideal, vars, w, irreds]} builds and loads a system of linear equations of weight \code{w} using the irreducible monomials \code{irreds} into \soft{FiniteFlow} to generate the companion matrices for each of the variables \code{vars} in the \code{ideal}.

\code{BuildCompanionMatrices[ideal, vars, \{wmin, wmax\}, irreds]} increases the seed iteratively from \code{wmin} until the system closes or \code{wmax} is reached.

\begin{center}
\begin{tabular}{|l|l|p{0.55\linewidth}|}
\hline
\multicolumn{3}{|c|}{\textbf{Options for} \code{BuildCompanionMatrices}} \\
\hline
\textbf{Option} & \textbf{Default Value} & \textbf{Description} \\
\hline
\texttt{"MonomialOrder"} & \texttt{Lexicographic} & monomial order to use \\
\texttt{"PrintDebugInfo"} & \texttt{0} & verbose printing with timings \\
\hline
\end{tabular}
\end{center}
\subsubsection{BuildTargetCompanionMatrices}
\code{BuildTargetCompanionMatrices[targets,cmatsystem]}
builds companion matrices for given target rational functions. \code{cmatsystem} should be the output of \code{BuildCompanionMatrices}. 

There are no options for this function.
\subsubsection{ReconstructTargetCompanionMatrices}
\code{ReconstructTargetCompanionMatrices[targetcmatsystem]}
reconstructs the remainder of rational functions encoded in the companion matrices. \code{targetcmatsystem} should be the output of \newline
\texttt{BuildTargetCompanionMatrices} or \code{BuildCompanionMatrices}.
\begin{center}
\begin{tabular}{|l|l|p{0.55\linewidth}|}
\hline
\multicolumn{3}{|c|}{\textbf{Options for} \code{ReconstructTargetCompanionMatrices}} \\
\hline
\textbf{Option} & \textbf{Default Value} & \textbf{Description} \\
\hline
\texttt{"cmat"} & \texttt{False} & reconstructs the full companion matrix \\
\texttt{"DeleteGraph"} & \texttt{True} & automatically deletes the \soft{FiniteFlow} graph after reconstruction \\
\texttt{"Vector"} & \texttt{False} & provide output already dotted with irreducible monomials or in vector form \\
\texttt{"PrintDebugInfo"} & \texttt{1} & prints sampling statistics from \soft{FiniteFlow} \\
\hline
\end{tabular}
\end{center}
\subsubsection{BuildCharacteristicPolynomials}
\code{BuildCharacteristicPolynomials[targetcmatsystem]}
Builds the characteristic polynomials for each companion matrix in \code{targetcmatsystem}.

\code{BuildCharacteristicPolynomials[targetcmatsystem,indexlist]}
builds the characteristic polynomials of the matrices indexed in \code{indexlist}. 

There are no options for this function.
\subsubsection{ReconstructCharacteristicPolynomials}
\code{ReconstructCharacteristicPolynomials[charpolys]}
reconstructs each coefficient of the characteristic polynomials produced by \code{BuildCharacteristicPolynomials}.

\code{ReconstructCharacteristicPolynomials[charpolys,coefficientlist]}
reconstructs only the terms given in \code{coefficientlist}.
\begin{center}
\begin{tabular}{|l|l|p{0.55\linewidth}|}
\hline
\multicolumn{3}{|c|}{\textbf{Options for} \code{ReconstructCharacteristicPolynomials}} \\
\hline
\textbf{Option} & \textbf{Default Value} & \textbf{Description} \\
\hline
\texttt{"PrintDebugInfo"} & \texttt{1} & prints sampling statistics from \soft{FiniteFlow} \\
\texttt{"DeleteGraph"} & \texttt{True} & automatically deletes the \soft{FiniteFlow} graph after reconstruction \\
\texttt{"Mod"} & \texttt{False} & reconstructs modulo a prime number \\
\texttt{"FFPrimeNo"} & \texttt{0} & if reconstructing modulo a prime, reconstructs modulo the specified \soft{FiniteFlow} prime \\
\hline
\end{tabular}
\end{center}
\subsubsection{FindEliminationMonomials}
\code{FindEliminationMonomials[ideal,\{x1,x2,...\},\{y1,y2,...\}]} computes the monomials appearing in the ideal where \code{\{x1,x2,..\}} have been eliminated using a numerical Gr\"obner Basis.

There are no options for this function.
\subsection{BuildEliminationSystems}
\code{BuildEliminationSystems[cmatsystem,monomials]} Builds and loads the equations required to eliminate variables from an ideal. \code{cmatsystem} and \code{monomials} should be the outputs of \code{BuildCompanionMatrices} and \code{FindEliminationMonomials} respectively.

There are no options for this function.
\subsection{ReconstructEliminationSystems}
\code{ReconstructEliminationSystems[elimSystem]} reconstructs each coefficient of the eliminated ideal. \code{elimSystem} should be the output of \code{BuildEliminationSystems}.

\begin{center}
\begin{tabular}{|l|l|p{0.55\linewidth}|}
\hline
\multicolumn{3}{|c|}{\textbf{Options for} \code{ReconstructEliminationSystems}} \\
\hline
\textbf{Option} & \textbf{Default Value} & \textbf{Description} \\
\hline
\texttt{"Vector"} & \texttt{False} & provide output already dotted with monomials in the eliminated ideal or in vector form \\
\texttt{"PrintDebugInfo"} & \texttt{1} & prints sampling statistics from \soft{FiniteFlow} \\
\texttt{"DeleteGraph"} & \texttt{True} & automatically deletes the \soft{FiniteFlow} graph after reconstruction \\
\texttt{"Mod"} & \texttt{False} & reconstructs modulo a prime number \\
\texttt{"FFPrimeNo"} & \texttt{0} & if reconstructing modulo a prime, reconstructs modulo the specified \soft{FiniteFlow} prime \\
\hline
\end{tabular}
\end{center}
\subsubsection{BuildPolynomialSystem}
\code{BuildPolynomialSystem[targets,ideal,vars,w]}
Builds and loads a system of linear equations to weight \code{w} into \soft{FiniteFlow} to polynomially reduce the targets with respect to the \code{ideal}.

\code{BuildPolynomialSystem[targets,ideal,vars,\{wmin,wmax\}]}
Increases the seed iteratively from \code{wmin} until the system closes or \code{wmax} is reached.
\begin{center}
\begin{tabular}{|l|l|p{0.50\linewidth}|}
\hline
\multicolumn{3}{|c|}{\textbf{Options for} \code{BuildPolynomialSystem}} \\
\hline
\textbf{Option} & \textbf{Default Value} & \textbf{Description} \\
\hline
\texttt{"MonomialOrder"} & \texttt{Lexicographic} & monomial order to use \\
\texttt{"IrreducibleMonomials"} & \texttt{\{\}} & check against provided monomials \\
\texttt{"EliminateVariables"} & \texttt{\{\{\},0\}} & elimination of variables from polynomial systems \\
\texttt{"PrintDebugInfo"} & \texttt{0} & verbose printing with timings \\
\hline
\end{tabular}
\end{center}
\subsubsection{ReconstructPolynomialRemainder}
\code{ReconstructPolynomialRemainder} takes the system generated by \code{BuildPolynomialSystem} and reconstructs the output of the polynomial division, namely the coefficients of the irreducible monomials.
\begin{center}
\begin{tabular}{|l|l|p{0.55\linewidth}|}
\hline
\multicolumn{3}{|c|}{\textbf{Options for} \code{ReconstructPolynomialRemainder}} \\
\hline
\textbf{Option} & \textbf{Default Value} & \textbf{Description} \\
\hline
\texttt{"Vector"} & \texttt{False} & provide output already dotted with irreducible monomials or in vector form \\
\texttt{"PrintDebugInfo"} & \texttt{1} & prints sampling statistics from \soft{FiniteFlow} \\
\texttt{"DeleteGraph"} & \texttt{True} & automatically deletes the \soft{FiniteFlow} graph after reconstruction \\
\hline
\end{tabular}
\end{center}
\subsubsection{SortVariables}
\code{SortVariables[ideal,vars]}
Sorts the \code{variables} to try make Gr\"obner Basis computations faster, based on \cite{Boege:1986, Lichtblau:1996}.
\begin{center}
\begin{tabular}{|l|l|p{0.23\linewidth}|}
\hline
\multicolumn{3}{|c|}{\textbf{Options for} \code{SortVariables}} \\
\hline
\textbf{Option} & \textbf{Default Value} & \textbf{Description} \\
\hline
\texttt{"MonomialOrder"} & \texttt{Lexicographic} & monomial order to use \\
\hline
\end{tabular}
\end{center}
\subsubsection{SPQRDet}
\code{SPQRDet[matrix]}
computes the determinant of a \code{matrix} using the Faddeev-LeVerrier algorithm.
\begin{center}
\begin{tabular}{|l|l|p{0.55\linewidth}|}
\hline
\multicolumn{3}{|c|}{\textbf{Options for} \code{SPQRDet}} \\
\hline
\textbf{Option} & \textbf{Default Value} & \textbf{Description} \\
\hline
\texttt{"PrintDebugInfo"} & \texttt{1} & prints sampling statistics from \soft{FiniteFlow} \\
\texttt{"Mod"} & \texttt{False} & reconstructs modulo a prime number \\
\texttt{"FFPrimeNo"} & \texttt{0} & if reconstructing modulo a prime, reconstructs modulo the specified \soft{FiniteFlow} prime \\
\hline
\end{tabular}
\end{center}

%% file: sections/examples.tex
\newpage
\section{Select Examples and Applications}\label{sec:examples}
In this section proof of concept examples and benchmarks are provided to showcase how \spqr{} can be applied to tackle state of the art problems, both in mathematics and in high energy physics. Specifically we consider the construction of Macaulay resultants as well as the determination of Landau singularities for Feynman integrals.
\subsection{Macaulay Resultants}
\subsubsection{Background}
Consider a polynomial ideal $\Ideal$ comprised of $\numvars$ variables (unknowns), $\numgens = \numvars + 1$ equations, as well as various parameters. In general, such a system is overdetermined and thus has no solutions. 

For special values of the parameters of the system however, the equations may no longer become overdetermined and a solution can exist. Such information is encoded in the \emph{Macaulay resultant} $\mathcal{R}$ \cite{Sturmfels:2002,CoxLittleOshea:2005,Cox:2015ode}. This object is a new polynomial which depends \textit{only} on the coefficients of $\Ideal$, which is defined to vanish precisely on the parameter configurations which allow for $\Ideal$ to have roots. For example, suppose
\begin{equation}
    \Ideal = \langle x-a,a\,x-1\rangle\,,
\end{equation}
where $x$ is the single variable and $a$ is a parameter. $V(\Ideal)\neq\emptyset$ only when $a = \pm 1$. Thus, we have
\begin{equation}
    \mathcal{R}(a) = (a-1)(a+1)\,.
\end{equation}
\subsubsection{Implementation in \spqr{}}\label{sec:resultant_implementation}
By their nature, the computation of Macaulay resultants inevitably involves dealing with ideals with multiple variables and parameters. Thus, as discussed in \cref{sec:advantages_and_disadvantages}, such computations often lend themselves well to \spqr{}'s finite fields approach. 

Suppose $\Ideal$ is comprised of $\numgens = \numvars + 1$ equations, $\numvars$ variables given by \{$x_1,\cdots,x_\numvars\}$, and $p$ parameters $\{a_1,\cdots,a_p\}$. To compute the Macaulay resultant with \spqr{}, one \emph{parameter}, say $a_p$, is ``promoted" to become a variable. This new system will now have $\numvars+1$ variables and $\numvars+1$ equations, and so generically will admit (zero-dimensional) solutions. The resultant can then be calculated by eliminating $\{x_1,\cdots, x_\numvars\}$ from this system, which can be done efficiently in \spqr{} by building the companion matrix $M_{a_p}$ and its respective characteristic polynomial.

By the nature of \spqr{}'s working ring, as discussed in \cref{sec:which_ring}, this approach may miss factors of the resultant that only depend on $\{a_1,\cdots a_{p-1}\}$. Nevertheless checking and correcting for such behaviour is straightforward: the output of \spqr{} can be compared against standard computer algebra procedures on a numerical slice to check if factors are missing. If any are found these can in turn be reconstructed by promoting the relevant parameters $a_i$ to variables instead. Since the missing terms cannot depend on $a_p$, the reconstruction can be performed on a partial numerical slice, which nearly always will result in a lighter computations compared to the first step.

Concretely, consider the ideal
\begin{equation}
    \Ideal = \langle a+x^2 y^2+y^3+z-1,a x+c x y^2+c y+z^2-2,a+b x y^2+b+x^2 y^2,-c+d x z+x y z+1 \rangle\,,
\end{equation}
in the original variables $\{x,y,z\}$ and parameters $\{a,b,c,d\}$. Suppose the task is to compute the macaulay resultant $\mathcal{R}(a,b,c,d)$ for $\Ideal$. We begin by adding $d$ to the list of variables, and (optionally) run the command \code{SortVariables} to attempt to find an optimal ordering:
\begin{lstlisting}
ideal = {
    -1 + a + x^2*y^2 + y^3 + z,
    -2 + a*x + c*y + c*x*y^2 + z^2,
    a + b + b*x*y^2 + x^2*y^2, 
    1 - c + d*x*z + x*y*z
 };
variables = SortVariables[ideal,{x,y,z,d}]
(*{d, z, x, y}*)
\end{lstlisting}
The rest of the process is identical to the elimination example already presented in \cref{sec:var_elim_w_cmat}: we find the irreducible monomials of this new ideal, and build the companion matrices for $\{x,y,z,d\}$:
\begin{lstlisting}
irreds = FindIrreducibleMonomials[ideal, variables, 
        "MonomialOrder" -> DegreeReverseLexicographic];
irreds // Length
(*14*)
cmats = BuildCompanionMatrices[ideal, variables, {1,10}, irreds, 
        "MonomialOrder" -> DegreeReverseLexicographic];
\end{lstlisting}
The characteristic polynomial for $M_d$ is then built and reconstructed:
\begin{lstlisting}
chard = BuildCharacteristicPolynomials[cmats,{1}];
res = ReconstructCharacteristicPolynomials[chard] // First;
\end{lstlisting}
Finally the resultant is formed by explicitly reintroducing $d$ and processing the result with \soft{Mathematica}'s built in tools:
\begin{lstlisting}
resultantSPQR = Power[d,Range[(irreds//Length)+1]-1] . res // Factor // Numerator;
resultantSPQR // Length
(*27062*)
resultantSPQR // ByteCount
(*9376024*)
\end{lstlisting}
This expression must now be checked against univariate numerical slices in $\{a,b,c\}$ to ensure that no factors have been missed. In \soft{Mathematica} this can be done with:
\begin{lstlisting}
(*check for a*)
ksub = {b,c,d} -> RandomInteger[10^10,3] // Thread;
expr1 = resultantSPQR // ReplaceAll[ksub];
expr2 = GroebnerBasis[ideal // ReplaceAll[ksub], Complement[{a,b,c,d}, ksub[[;;,1]]], {x,y,z},
        CoefficientDomain -> RationalFunctions] // First;
expr1 / expr2 // Factor // Variables
(*{}*)
\end{lstlisting}
\begin{minipage}{\linewidth}
\begin{lstlisting}
(*check for b*)
ksub = {a,c,d} -> RandomInteger[10^10,3] // Thread;
expr1 = resultantSPQR // ReplaceAll[ksub];
expr2 = GroebnerBasis[ideal // ReplaceAll[ksub], Complement[{a,b,c,d}, ksub[[;;,1]]], {x,y,z},
        CoefficientDomain -> RationalFunctions] // First;
expr1 / expr2 // Factor // Variables
(*{}*)
\end{lstlisting}
\end{minipage}
\begin{lstlisting}
(*check for c*)
ksub = {a,b,d} -> RandomInteger[10^10,3] // Thread;
expr1 = resultantSPQR // ReplaceAll[ksub];
expr2 = GroebnerBasis[ideal // ReplaceAll[ksub], Complement[{a,b,c,d}, ksub[[;;,1]]], {x,y,z},
        CoefficientDomain -> RationalFunctions] // First;
expr1 / expr2 // Factor // Variables
(*{}*)
\end{lstlisting}
Since all numerical slices agree, no extra factors are missing and no further analysis is required.
\subsubsection{Benchmark}\label{sec:benchmark_resultant}
The qualitative observations discussed in \cref{sec:advantages_and_disadvantages} can be made more concrete by computing $\mathcal{R}(a,b,c,d)$ from \cref{sec:resultant_implementation} with various computer algebra systems. We compare the performance of \spqr{} against \soft{Singular} \cite{DGPS}, \soft{Macaulay2} \cite{M2} and \soft{msolve} \cite{msolve} on various numerical slices of $\Ideal$: by substituting some parameters to numerical values, it is possible to vary the ``parameter complexity" of $\Ideal$, whilst keeping the ``variable complexity" constant.

It is important to note explicitly that a direct comparison between \spqr{} and other computer algebra approaches is difficult: finite field reconstruction algorithms are by nature probabilistic, and thus depending on the desired rigour, could be ruled out from the start. Furthermore, the sampling and reconstruction phase in \spqr{} is heavily multithreaded, which may be advantageous or disadvantageous depending on the configuration at hand. Finally, specifically to the elimination problem being considered, we adopt \spqr{}'s companion matrix approach to eliminating variables, which is a different strategy to the default block elimination ordering approach used by \soft{Singular} \soft{macaulay2} and \soft{msolve}.

From \cref{tab:timings_res}
\begin{table}[h!]
\centering
\begin{tabular}{
  l
  r@{\hspace{1em}}r
  r@{\hspace{1em}}r
  r@{\hspace{1em}}r
  r@{\hspace{1em}}r
}
\toprule
 & \multicolumn{2}{c}{\soft{Singular}}
 & \multicolumn{2}{c}{\soft{Macaulay2}}
 & \multicolumn{2}{c}{\soft{msolve}}
 & \multicolumn{2}{c}{\spqr{}} \\
\cmidrule(lr){2-3} \cmidrule(lr){4-5} \cmidrule(lr){6-7} \cmidrule(lr){8-9}
\textbf{Resultant}
 & \multicolumn{1}{c}{Time} & \multicolumn{1}{c}{RAM}
 & \multicolumn{1}{c}{Time} & \multicolumn{1}{c}{RAM}
 & \multicolumn{1}{c}{Time} & \multicolumn{1}{c}{RAM}
 & \multicolumn{1}{c}{Time} & \multicolumn{1}{c}{RAM} \\
\midrule
$\mathcal{R}(3,5,7,d)$
  & 0.01 s  & 11 MB
  & 0.06 s  & 101 MB
  & 0.003 s & 9.5 MB
  & 0.34 s  & 182 MB \\
$\mathcal{R}(3,5,c,d)$
  & 53.29 s & 26 MB
  & 12.40 s & 106 MB
  & 0.10 s  & 14.5 MB
  & 0.58 s  & 224 MB \\
$\mathcal{R}(3,b,c,d)$
  & $>$14 d   & $>$71 GB
  & $>$14 d   & $>$149 GB
  & 3065 s  & 23 GB
  & 1.06 s  & 0.27 GB \\
$\mathcal{R}(a,b,c,d)$
  & ? d       & ? GB
  & ? d       & ? GB
  & $>$4 d  & $>$1 TB
  & 3.09 s  & 0.29 GB \\
\bottomrule
\end{tabular}
\caption{Computation times for $\mathcal{R}(a,b,c,d)$ on various numerical slices across different systems. \soft{Singular} and \soft{Macaulay} did not finish after fourteen days of computation time on $\mathcal{R}(3,b,c,d)$, and were thus not attempted on the full resultant. \soft{msolve} after four days on $\mathcal{R}(a,b,c,d)$ exceeded the maximum available memory and thus did not terminate. Evaluations performed on a 2 x AMD EPYC 7532 32-Core Processor with 1 TB of RAM.}
\label{tab:timings_res}
\end{table}
it is clear that on the fully numerical slice $\mathcal{R}(3,5,7,d)$, many computer algebra implementations outperform \spqr{} by orders of magnitude. This is due to the large overhead in building and solving the relevant Macaulay system inside \spqr{}. Furthermore, there are no parameters to reconstruct, rendering \spqr{}'s finite fields and reconstruction pipeline mostly superfluous. Indeed, this numerical slice serves as an example with ``variable complexity" but with no ``parameter complexity". Furthermore, based on the computation times it is clear that the ``variable complexity" of this problem is comfortably within reach of all modern computer algebra systems. For a more complicated examples where the variables complexity alone challenges traditional algorithms, we expect the gap with \spqr{} to be further increased.

On smaller numerical slices however the computation times change drastically. The symbolic approaches scale poorly as parameters are reintroduced, which we suspect is due to intermediate expression swell. This severely impacts both the computation time and memory required. The finite field approach implemented in \spqr{} does not suffer from this problem and thus seems to scale better. 

Indeed, \soft{Singular} and \soft{Macaulay2} did not terminate on the slice $\mathcal{R}(3,b,c,d)$ with over $14$ days of compute time, and were thus not attempted on the full resultant. \soft{msolve} did manage to compute $\mathcal{R}(3,b,c,d)$, but exceeded the maximum 1 TB of RAM available after 4 days when computing $\mathcal{R}(a,b,c,d)$. Thus, with the full parameter dependence, \spqr{} results in at least 5-6 orders of magnitude improvement in compute time, and 3-4 in memory usage.

To build \cref{tab:timings_res} the following \soft{Singular} \soft{Macaulay2} and \soft{msolve} codes were respectively used:
\begin{lstlisting}
ring R = 0, (x,y,z,a,b,c,d),dp;
ideal I = -1 + a + x^2*y^2 + y^3 + z, -2 + a*x + c*y + c*x*y^2 + z^2,
           a + b + b*x*y^2 + x^2*y^2, 1 - c + d*x*z + x*y*z;
timer=0;
system("--ticks-per-sec",1000);
int t=timer;
ideal J = eliminate(I,x*y*z);
timer-t;
\end{lstlisting}
\begin{lstlisting}
R = QQ[x,y,z,a,b,c,d]
L = {-1 + a + x^2*y^2 + y^3 + z, -2 + a*x + c*y + c*x*y^2 + z^2,
    a + b + b*x*y^2 + x^2*y^2, 1 - c + d*x*z + x*y*z}
I = ideal(L)
timing(eliminate({x,y,z},I);)
\end{lstlisting}
\begin{lstlisting}
msolve -e 3 -g 2 -f benchmark -o out
\end{lstlisting}
with the file \code{benchmark} containing
\begin{lstlisting}
x,y,z,d,c,b,a
0
-1+a+x^2*y^2+y^3+z, -2+a*x+c*y+c*x*y^2+z^2, a+b+b*x*y^2+x^2*y^2, 1-c+d*x*z+x*y*z
\end{lstlisting}
For \spqr{} the timings refer to the all the code presented in \cref{sec:resultant_implementation}, including the verification steps performed with \soft{Mathematica}. The full form of $\mathcal{R}(a,b,c,d)$ can be found in the tutorial installed along with the \spqr{} package. 
\subsection{Landau Analysis}
The specific example considered in \cref{sec:resultant_implementation,sec:benchmark_resultant} was purely illustrative and of limited practical interest. In this section we instead focus on how \spqr{} can be used to tackle a class of polynomial ideals motivated by high energy physics.
\subsubsection{Background}\label{sec:landau_background}
Feynman integrals are ubiquitous in modern high energy physics calculations. Through the use of a parametric representation (see \cite{Weinzierl:2022eaz} for a review), mathematically they can be interpreted as \textit{twisted period} or \textit{Euler} integrals, which can be chosen to take the form \cite{Lee:2013hzt},
\begin{equation}
    I(\svec) \sim \int_0^\infty \left(\mathcal{U}(\xvec)+\mathcal{F}(\xvec, \svec)\right)^{-d/2}\, \frac{\dd x_1}{x_1} \wedge \cdots \wedge \frac{\dd x_n}{x_n}\,,
\end{equation}
where an irrelevant prefactor for this discussion has been omitted. $\mathcal{U}$ and $\mathcal{F}$ are polynomials depending both on a set of integration (Schwinger) variables $\xvec$ and a set of (Mandelstam/kinematic) parameters $\svec$, on which $I$ ultimately depends. The parameter $d$ is taken to be generic, and thus $\mathcal{G}^{-d/2}$ is a multivalued function (twist) with branch points at the roots of $\mathcal{U}+\mathcal{F}$.

In general $I(\svec)$ is a complicated expression with an involved branch cut structure. When attempting to evaluate Feynman integrals, knowledge of the branch points can be of great help, in particular for the construction of the differential equation systems they obey \cite{Gehrmann:1999as,Remiddi:1997ny}. The methods and algorithms to obtain this information are collectively known as \emph{Landau Analysis} \cite{Bjorken:1959fd,Landau:1959fi,Nakanishi:1959jzx,Cutkosky:1960sp}.

Specifically, given a specific $\mathcal{U}(\xvec)\,,\mathcal{F}(\xvec,\svec)$ the goal of most Landau algorithms is to compute the \emph{Landau variety} $l(\svec)$. Similar to a resultant, this is a polynomial in the kinematics $\svec$, such that for $l(\svec^*)=0$, $\svec^*$ is a branch point of $I$. In practice $l(\svec)$ often factors into smaller polynomials $l(\svec) = l_1(\svec)\cdots l_k(\svec)$. Each irreducible polynomial $l_i(\svec)$ is known as a \emph{Landau singularity}\footnote{For reviews of Landau analysis we refer the reader to \cite{Weinzierl:2022eaz,Badger:2023eqz}.}.

In recent years there has been an enormous growth of interest and progress in computing Landau singularities, spurred by both theoretical and technological breakthroughs \cite{Dlapa:2023cvx,Brown:2009ta,Panzer:2014caa,Klausen:2021yrt,Hannesdottir:2022xki,Correia:2021etg,Mizera:2021icv,Berghoff:2022mqu,Fevola:2023kaw,Fevola:2023fzn,Helmer:2024wax,Correia:2025yao,Caron-Huot:2024brh}.

The traditional and most used approach to computing Landau singularities is via the \emph{Landau equations} \cite{Landau:1959fi}: one formulation seeks values of $\svec$ such that the equation system
\begin{equation}\label{eq:landau}
	\mathcal{F} = 0\,, \quad x_i\, \frac{\partial \mathcal{F}}{\partial x_i} = 0 \quad \forall\, i\,.
\end{equation}
has solutions. In practice to find \emph{all} the Landau singularities for a Feynman diagram, different equation systems along with blow ups may need to be considered \cite{Fevola:2023fzn}. Nevertheless even after such manipulations the end result is always a set of polynomial equations which need to be solved for.
\subsubsection{Implementation in \spqr{}}\label{sec:landau_implementation}
Solving \cref{eq:landau} amounts to eliminating variables from polynomial systems with many parameters, and thus lends itself well to \spqr{}'s approach. One approach to finding Landau singularities with \spqr{} is as follows: we consider the ideal
\begin{equation}
	\Ideal = \left\langle \mathcal{F}, \frac{\partial \mathcal{F}}{\partial x_1}, \cdots, \frac{\partial \mathcal{F}}{\partial x_\numvars}, 1-x_0 (x_1 \cdots x_n) \right\rangle\,,
\end{equation}
where compared to \cref{eq:landau} the simpler solutions involving $x_i = 0$ are excluded, as they result in simpler (subsector) subsystems which can be solved separately.

As in \cref{sec:resultant_implementation}, $\Ideal$ is overdetermined, as there are $\numvars+2$ equations and $\numvars+1$ unknowns. Thus finding a set of Landau singularities can be translated to finding the multivariate resultant of $\Ideal$. In \spqr{} one thus proceeds by again ``promoting" one parameter in $\svec$, say $s_1$ to a variable, and eliminating the $\numvars +1$ variables $\{x_0,\cdots x_\numvars\}$ from the resulting $\numvars+2$ variable system.

As before this approach will miss singularities that depend only on $\{s_2,\cdots\}$. Exactly as in \cref{sec:benchmark_resultant}, this can once again be checked for by comparing against standard computer algebra procedures on a numerical slice. If any factors are missing these can in turn be reconstructed by promoting the relevant parameter to become a variable instead, before a simpler reconstruction on a partial numerical slice. 
\subsubsection{Benchmark}
This method can be applied to state of the art diagrams for which sets of Landau singularities have already been studied \cite{Fevola:2023fzn,Correia:2025yao}. Concretely, we consider the diagram \texttt{env-equal-zero} for which a set of already computed singularities can be found at \cite{PLD_website,StrangeQuark007_SOFIA_PLDdatabase_2025}. Its respective $\mathcal{F}$ polynomial contains 6 variables $\xvec=\{x_1,\cdots,x_6\}$ as well as 3 parameters $\svec = \{m^2,s,t\}$.

In \soft{Mathematica} the setup is given by:
\begin{lstlisting}
f = {m2 x1^2 x2 x3 + m2 x1 x2^2 x3 + m2 x1 x2 x3^2 + m2 x1^2 x2 x4 + m2 x1 x2^2 x4 
	+ m2 x1^2 x3 x4 + 4 m2 x1 x2 x3 x4 - t x1 x2 x3 x4 + m2 x2^2 x3 x4 + m2 x1 x3^2 x4 
	+ m2 x2 x3^2 x4 + m2 x1 x2 x4^2 + m2 x1 x3 x4^2 + m2 x2 x3 x4^2 + m2 x1^2 x2 x5 
	+ m2 x1 x2^2 x5 + m2 x1^2 x3 x5 + 3 m2 x1 x2 x3 x5 + m2 x1 x3^2 x5 + 3 m2 x1 x2 x4 x5 
	+ m2 x2^2 x4 x5 + 3 m2 x1 x3 x4 x5 + 3 m2 x2 x3 x4 x5 + m2 x3^2 x4 x5 + m2 x2 x4^2 x5
	+ m2 x3 x4^2 x5 + m2 x1 x2 x5^2 + m2 x1 x3 x5^2 + m2 x2 x4 x5^2 + m2 x3 x4 x5^2 
	+ m2 x1^2 x3 x6 + 3 m2 x1 x2 x3 x6 + m2 x2^2 x3 x6 + m2 x1 x3^2 x6 + m2 x2 x3^2 x6 
	+ m2 x1^2 x4 x6 + 3 m2 x1 x2 x4 x6 + m2 x2^2 x4 x6 + 3 m2 x1 x3 x4 x6 + 3 m2 x2 x3 x4 x6 
	+ m2 x1 x4^2 x6 + m2 x2 x4^2 x6 + m2 x1^2 x5 x6 + 3 m2 x1 x2 x5 x6 + m2 x2^2 x5 x6 
	+ 4 m2 x1 x3 x5 x6 - s x1 x3 x5 x6 + 3 m2 x2 x3 x5 x6 + m2 x3^2 x5 x6 + 3 m2 x1 x4 x5 x6 
	+ 4 m2 x2 x4 x5 x6 + s x2 x4 x5 x6 + t x2 x4 x5 x6 + 3 m2 x3 x4 x5 x6 + m2 x4^2 x5 x6 
	+ m2 x1 x5^2 x6 + m2 x2 x5^2 x6 + m2 x3 x5^2 x6 + m2 x4 x5^2 x6 + m2 x1 x3 x6^2 
	+ m2 x2 x3 x6^2 + m2 x1 x4 x6^2 + m2 x2 x4 x6^2 + m2 x1 x5 x6^2 + m2 x2 x5 x6^2 
	+ m2 x3 x5 x6^2 + m2 x4 x5 x6^2} // First;
(*Landau singularities are homogenous*)
ksub = {t->1};
ideal = Join[{f}, D[f, {{x1, x2, x3, x4, x5, x6}}], {1 - x0*x1*x2*x3*x4*x5*x6}] /. ksub;
vars = {m2, x0, x1, x2, x3, x4, x5, x6};
\end{lstlisting}
where $m^2$ has already been ``promoted" to a variable inside $\code{vars}$ and we have set $t=1$ to exploit the homogeneity of Landau singularities. We now turn to eliminating $\{x_0,\cdots x_6\}$ in this system. An important difference with the example in \cref{sec:resultant_implementation} is that this \code{ideal} is \emph{not} zero-dimensional, as can be verified with
\begin{lstlisting}
FindIrreducibleMonomials[ideal, vars, "MonomialOrder" ->DegreeReverseLexicographic]
(*\[Infinity]*)
\end{lstlisting}
Since companion matrices require zero-dimensional systems, they cannot be used here. One can proceed with a ``trick" to restore zero-dimensionality: since landau singularities cannot depend on any of the variables $\xvec$, the $m^2$ coordinates of $V(\Ideal)$ must be point-like. Thus, one can intersect the solution space of \code{ideal} with sufficiently generic hyperplanes until a zero-dimensional solution space is reached, the roots of which projected onto $m^2$ will remain unchanged\footnote{An alternative to this approach in \spqr{} is to forego zero-dimensional systems and companion matrices to instead use the more traditional approach of elimination orders.}.

For this example, intersecting with the hyperplane defined by $x_6 = \text{constant}$ suffices to restore zero-dimensionality. This linear condition can be substituted inside $\code{ideal}$ to obtain
\begin{lstlisting}
ideal0dim = ideal // ReplaceAll[x6 -> RandomInteger[{1, 10^15}]];
vars0dim = vars[[1 ;; -2]];
irreds = FindIrreducibleMonomials[ideal0dim, vars0dim, 
		"MonomialOrder" ->DegreeReverseLexicographic];
irreds // Length
(*48*)
\end{lstlisting}
Companion matrices can now be built:
\begin{lstlisting}
cmats = BuildCompanionMatrices[ideal0dim, vars0dim, {11, 15}, irreds, 
        "MonomialOrder"->DegreeReverseLexicographic, "PrintDebugInfo"->2];
\end{lstlisting}
and used to eliminate $\{x_0,\cdots, x_5\}$ from this system. We proceed using the method described in \cref{sec:elim_cmat_ansatz}:
\begin{lstlisting}
elimMons = FindEliminationMonomials[ideal0dim, {x0, x1, x2, x3, x4, x5}, {m2}];
elimSyst = BuildEliminationSystems[cmats,elimMons];
elim = ReconstructEliminationSystems[elimSyst];
\end{lstlisting}
The Landau singularities are now the factors of \code{elim}. They can be recovered with:
\begin{lstlisting}
elimNumerator = elim // First // Together // Numerator;
landauinhomog = elimNumerator // FactorList // Flatten // DeleteCases[x_ /; IntegerQ[x]];
landau = landauinhomog // Map[ResourceFunction["PolynomialHomogenize"][#,{s,m2},t]&] // Sort;
landau // Length
(*5*)
\end{lstlisting}
where after separating each factor the $t$ dependence is restored by homogenising. Out of the 5 singularities contained in \code{landau}, this approach reproduces the previously most known complicated letter, $\code{landau[[1]]}=27 (m^2)^3 + 4 s^2 t + 4 s t^2$. The remaining four letters are significantly more involved and represent new previously unknown singularities for this Feynman integral. The full set of these new singularities is given in the examples section of \spqr{}'s tutorial. 

Compared to the example discussed in \cref{sec:benchmark_resultant} this computation is more challenging, taking approximately $30$ minutes and roughly $20$GB of memory on the same machine as in \cref{tab:timings_res}.

By computing the Euler characteristic \cite{Fevola:2023fzn,Chestnov:2023kww} via critical point counting \cite{Lee:2013hzt,Mastrolia:2018uzb,Frellesvig:2019uqt} implemented in \cite{Crisanti:2025gcs}, one can verify that these new singularities are not spurious. Indeed the Euler characteristic drops from a generic value of $\chi = 181$ to $\chi = \{177,179,179,179,180\}$ when restricted to each entry of \code{landau} respectively.

We also expect that \spqr{}'s elimination routines could prove helpful in other Landau analysis methods that do not involve solving the Landau equations directly, such as the Whitney stratification approach presented in \cite{Helmer:2024wax}.

%% file: sections/conclusions.tex
\newpage
\section{Conclusions and Outlook}\label{sec:conclusions}

The study of polynomial systems is ubiquitous in mathematics, physics and beyond. In this work we presented \spqr{}, a new \soft{Mathematica} package for tackling division and elimination problems in systems of polynomial equations.

\spqr{} processes polynomial systems differently to many other programs: all algorithms are systematically recast as solving linear systems of equations and matrix algebra. The key innovation is that these operations are then in turn implemented in terms of finite field sampling and black box reconstruction pipelines.

Essentially \spqr{} thus borrows many of the techniques that have pushed the state of the art in scattering amplitude computations, and repackages them to problems in polynomial algebra. This allows \spqr{} to effectively avoid large intermediate expressions, which can significantly impact analytic computer algebra approaches.

Indeed we argue that expression swell can often be the bottleneck in processing polynomial systems, in many cases more so than the efficiency of Gr\"obner basis algorithm itself. \spqr{}'s approach is thus best suited to ideals with complicated parameter (coefficient domain) dependency, but moderate variable complexity. For systems of this kind we find \spqr{}'s approach to be extremely effective, significantly pushing the state of the art when compared to other publicly available implementations.

To this end we benchmarked the package’s elimination tools on state of the art Macaulay resultant computations. Across these tests, \spqr{} delivered marked improvements over other computer algebra systems, reducing both runtime and memory by at least 5-6 and 3-4 orders of magnitude respectively, when compared to \soft{Singular}, \soft{Macaulay2} and \soft{msolve}. We also tested \spqr{} on ideals motivated by theoretical physics: when applied to previously studied state of the art Feynman integrals, \spqr{} was able to find new previously missed Landau singularities without requiring a specialised solver specific to this task.

Despite its extensive usage of finite field sampling and reconstruction, the package has been designed to require no knowledge from the end user of its inner workings. At the same time however it remains flexible enough to fit into a wide range of algorithmic pipelines.

The ideas behind \spqr{} open several directions for future exploration. \spqr{} as of current builds (numerical) Gr\"obner bases via solving Macaulay systems of equations. Whilst this is similar to modern dedicated algorithms such as \soft{F4}, \spqr{}'s approach is not as fine tuned. Integrating a fast numerical Gr\"obner basis implementation such as \soft{msolve} into \spqr{} would help push the package's scope even further.

Another potential improvement could be made in the reconstruction phase of \spqr{}'s pipeline. In particular, \spqr{}'s back end \soft{FiniteFlow} reconstructs all expressions fully expanded. An algorithm that attempts to understand the factorisation structure of the output could in many cases save several orders of magnitude in sample points. We expect such a procedure to be particularly beneficial for the reconstruction of Landau singularities, as we have often observed strong factorisation of the output in such cases.

Finally, we hope that the ideas behind \spqr{} will enjoy broad application in other scientific fields, beyond just the physics and mathematics motivated examples primarily presented in this work.